\documentclass[prd,twocolumn,floatfix,preprintnumbers,reprint,showpacs]{revtex4}
\usepackage{epsfig,amsmath,amssymb,bm,color,graphicx}
\usepackage{afterpage}
\usepackage{pdfpages}

\usepackage{color}

\newcommand{\bfx}{\mathbf{x}}

\newcommand{\be}{\begin{equation}}
\newcommand{\ee}{\end{equation}}

\usepackage{color}

\begin{document}

\title{Early Structure Formation Constraints on the Ultra-Light Axion in the Post-Inflation Scenario}
\author{Vid Ir\v{s}i\v{c}$^{1,2,3}$}\thanks{E-mail: vi223@cam.ac.uk (VI)}
\author{Huangyu Xiao$^{4}$}
\author{Matthew McQuinn$^{1}$}

\smallskip
\affiliation{
$^{1}$University of Washington, Department of Astronomy, 3910 15th Ave
NE, WA 98195-1580 Seattle, USA\\
$^{2}$Kavli Institute for Cosmology, University of Cambridge, Madingley Road, Cambridge CB3 0HA, UK\\
$^{3}$Cavendish Laboratory, University of Cambridge, 19 J. J. Thomson Ave., Cambridge CB3 0HE, UK\\
$^{4}$University of Washington, Department of Physics, 3910 15th Ave
NE, WA 98195-1580 Seattle, USA\\
}

\begin{abstract}
Many works have concentrated on the observable signatures of the dark matter being an ultralight axion-like particle (ALP).  We concentrate on a particularly dramatic signature in the late-time cosmological matter power spectrum that occurs if the symmetry breaking that establishes the ALP happens after inflation -- white-noise density fluctuations that dominate at small scales over the adiabatic fluctuations from inflation.  These fluctuations alter the early history of nonlinear structure formation.  We find that for symmetry breaking scales of $f_A \sim 10^{13}-10^{15}$GeV, which requires a high effective maximum temperature after inflation, ALP dark matter with particle mass of $m_A \sim 10^{-13}-10^{-20}$eV could significantly change the number of high-redshift dwarf galaxies, the reionization history, and the Ly$\alpha$ forest.  We consider all three observables.  We find that the Ly$\alpha$ forest is the most constraining of current observables, excluding $f_A \gtrsim 10^{15}$GeV ($m_A \lesssim 10^{-17}$eV) in the simplest model for the ALP and considerably lower values in models coupled to a hidden asymptotically-free strongly interacting sector ($f_A \gtrsim  10^{13}$GeV and $m_A \lesssim  10^{-13}$eV).  Observations that constrain the extremely high-redshift tail of reionization may disfavor similar levels of isocurvature fluctuations as the forest.   Future $z\sim 20-30$ 21cm observations have the potential to improve these constraints further using that the supersonic motions of the isocurvature-enhanced abundance of $\sim10^4M_\odot$ halos would shock heat the baryons, sourcing large BAO features.
\end{abstract}

\maketitle
\section{Introduction}

The nature of the dark matter remains one of the biggest unsolved puzzles in particle physics and cosmology.  We think that the dark matter is a particle produced in the early universe via one of several established mechanisms. The foremost has it thermally produced and its abundance freezing out when non-relativistic, which can result in the observed dark matter density if it has a weak-scale mass and interaction cross section -- the so-called `WIMP miracle' \citep[e.g.][]{1996PhR...267..195J}. After decades of searching for the WIMP, the limits on this scenario are becoming more stringent. Perhaps our second most favored mechanism is the misalignment mechanism, discovered for the axion of quantum chromodynamics \citep[QCD,][]{PhysRevLett.40.223, 1978PhRvL..40..279W, 1990eaun.book.....K}. At early times when Hubble rate is greater than axion mass -- a mass that is acquired by non-perturbative effects such as instantons--, the axion field is stuck outside of the minimum of its potential. However, when the Hubble rate later becomes smaller than axion mass, the axion field begins to oscillate coherently, behaving like non-relativistic matter with energy density set by its initial potential energy \citep{Preskill1982,Abbott1982,Dine1982,2016PhR...643....1M}.

The misalignment mechanism is also how the early universe could create dark matter in the form of ultra-light axion-like particles (ALPs; also known as fuzzy dark matter).
The misalignment mechanism may naturally produce an ALP relic abundance of order the dark matter abundance if the ALP is the Goldstone Boson arising from a broken GUT to Planck scale symmetry and if it later acquires a mass of $m_A \sim 10^{-20}$eV \citep{2017PhRvD..95d3541H}.  The non-perturbative mass generation can also naturally explain such ultralight masses, with $m_A \sim 10^{-20}$eV motivated by the estimated size of non-perturbative effects for the GUT coupling constant \citep{2016PhR...643....1M}.

Our study focuses on such ultra-light ALPs in the limit where the Peccei-Quinn symmetry breaking that establishes this particle (re)occurs after inflation.  For string theory-motivated models, the anticipated ranges for the symmetry breaking scale, $f_A$, are GUT to Planck scales \citep{2016PhR...643....1M, 2017PhRvD..95d3541H}, although models that allow a lower scale have been devised \citep{2006JHEP...06..051S}.  Too low of a symmetry breaking scale would not generate the dark matter abundance:  As our constraints probe, $m_A = 10^{-16} -10^{-20}$eV, this requires $f_A$ just below the GUT scale with $\sim 10^{15} -10^{16}~$GeV to generate the relic abundance.  These high values for $f_A$ (which are far above the Hubble scale during inflation so that this symmetry must be broken during this epoch) may be strained by CMB B-mode observations, which limit the energy scale of inflation to $V(\phi) \lesssim 1.7\times 10^{16}$GeV \citep{2018arXiv180706211P}.  Our mechanism requires the symmetry to be re-established after inflation.  This reestablishment can occur if the maximum post-inflation thermalization temperature is greater than $f_A$ \footnote{The maximum temperature is larger (in some models by orders of magnitude) than the reheat temperature \citep[e.g.][]{2003PhRvD..68l3505K}.} or instead during preheating where larger effective temperatures can naturally arise from the non-thermal distribution of resonantly produced particles \citep{1996PhLB..376...35T, 1996PhRvL..76.1011K}.  

We further consider models with an asymptotically-free strongly interacting sector that mimics the behavior of the QCD axion (in which the particle mass increases after the ALP behaves behaves like dark matter).  Such models allow a somewhat lower $f_A$ to match the dark matter abundance (down to $f_A \sim 10^{13}$GeV), at the cost of introducing a sub-MeV confinement scale.  The cosmological constant problem can be solved by hundreds of ALPs connected with strongly coupled sectors (as such sectors allow non-degenerate vacuum minima owing to higher instanton contributions), possibly with several hidden sectors per decade in energy \citep{2010PhRvD..81l3530A}. (See this endnote \footnote{In this strongly interacting `axiverse' scenario, any post-inflation ALP likely cannot have multiple non-degenerate vaccua to avoid a domain wall catastrophe.  Thus, the ALPs with non-degenerate vaccua would come into existence before inflation and have a small misalightment angle coherent over the cosmological volume so that they do not overclose the Universe, which perhaps could occur because of the anthropic principle \cite{2004hep.ph....8167W}.  For our results to apply of course, the ALPs that dominate the dark matter density would have to come into existence after inflation.} for more discussion of the strongly interacting `Axiverse' scenario, as there are some challenges to this scenario in our post-inflationary picture.)

Just like with the QCD axion in this post inflation limit, different causally disconnected patches will acquire different energy vaccua depending on the random angle $\theta \in [-\pi, \pi]$ the field rolled to after symmetry breaking in a given patch, with the Horizon scale setting the coherence length until $m_A \sim H$ \cite{KIBBLE, 1990eaun.book.....K}.  At this time, the vacuum energy is then converted into non-relativistic axions with number density $\propto \theta^2$, leading to order unity fluctuations in the abundance of axions on the horizon scale when $m_A \sim H$ \citep{1988PhLB..205..228H}.  The lighter the axion, the later this occurs, the larger the horizon-scale coherence length of the fluctuations. 

These isocurvature perturbations are potentially observable.  For the QCD axion \citep{1983PhLB..120..127P}, the mass contained in the horizon $M_{H(m_A)}$ when $m_A \sim H$ -- which is also the scale where there are order unity density fluctuations -- is $M_{H(m_A)} \sim 10^{-10}M_\odot $ \citep{Efstathiou1986,1988PhLB..205..228H, 2019JCAP...04..012V} (and axion self interactions can lead to larger enhancements on even smaller scales; \citep{1994PhRvD..49.5040K}).  This leads to the collapse of `axion miniclusters' near this mass scale at matter radiation equality, resulting in much denser dark matter structures than would be produced by the scale-invariant potential fluctuations from inflation.  Still, there is no smoking gun observable for verifying whether these minute structures exist, although see \cite{2019arXiv190801773D} for a promising possibility.  In contrast, for ultra-light axions that are relevant for small-scale structure problems, $M_{H(m_A)}$ can approach the sizes of dwarf galaxies, and the RMS fluctuations produced via these isocurvature fluctuations scale as $M^{-1}$, where $M$ is the average mass contained within a spherical volume.  These fluctuations are still larger than the inflationary perturbations even on mass scales of $M \gg M_{H(m_A)}$.  This property has been used to place constraints on the ultralight ALPs via the cosmic microwave background \citep[CMB;][]{2013PhRvD..87l1701M, Feix:2019lpo}.

\begin{figure*}
\includegraphics[width=13cm]{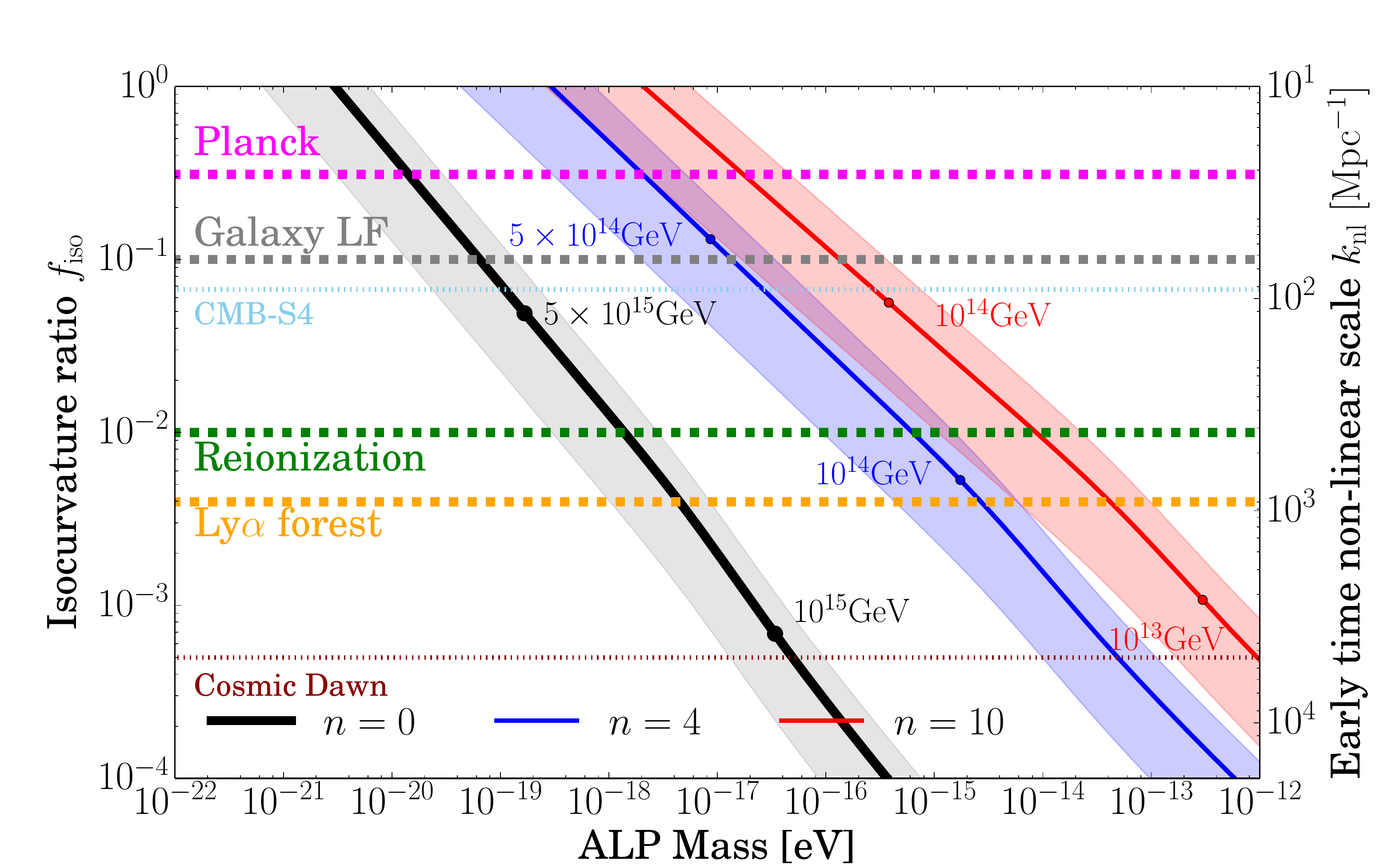}
\caption{The $f_{\rm iso}$ -- $m_a$ constraints for the ALP dark matter in the post-inflation scenario, assuming the ALP is all of the dark matter. The isocurvature to adiabatic ratio $f_{\rm iso}$ applies at the pivot scale of $k_\star = 0.05\;\mathrm{Mpc^{-1}}$, and the second $y$-axis gives the nonlinear scale defined by $\Delta_{\rm iso}^2 \equiv (k/k_{\rm nl})^3$ at early times.  The solid lines show this mapping for our fiducial choice of $A_{\rm osc} = 0.1$ and the shaded bands span $0.01 \leq A_{\rm osc} \leq0.3$. The horizontal dashed lines correspond to current upper limits on $f_{\rm iso}$ obtained using different data sets: the Planck 2018 CMB measurements in magenta from \cite{Feix:2019lpo}, Hubble Space Telescope galaxy luminosity function measurements in grey (\S~\ref{glf_sec}), a combination of constraints on the reionization history in green (\S~\ref{sfr_sec}), and finally Ly-$\alpha$ forest in orange (\S~\ref{lya_sec}). The horizontal dotted line in light blue is a forecast for CMB-S4  \cite{Feix:2019lpo}, while the horizontal dotted line in dark red is a rough forecast for where shock-heating should qualitatively change the $z\sim20$ 21cm signal (\S~\ref{cmb_sec}). The labeled dots give the value of the symmetry breaking scale $f_a$.
}
\label{fig_4}\label{fig_constraints}
\end{figure*}

This paper shows that other observables are much more constraining than the CMB.  We first focus on the the Ly$\alpha$ forest, which is the quasi-linear `large-scale' structure formation probe sensitive to the smallest scales.  In addition, we show that such isocurvature perturbations could significantly affect the formation of the first stars and galaxies in the redshift of $z\sim 6-20$ Universe, and discuss potential constraints.  Since these ioscurvature perturbations lead to the formation of dark matter halos at much higher redshifts than would occur in the standard cosmology, we also consider whether the shocks from these supersonic dense structures could ionize and heat the post-recombination universe.  Figure~\ref{fig_4} summarizes our constraints on the fractional amplitude of isocurvature fluctuations $f_{\rm iso}$ (defined shortly) and axion mass $m_a$, where dashed lines represent existing constraints and dotted represent forecasts for future efforts.

This paper is organized as follows.  Section \ref{sec:iso} describes the character of ALP isocurvature fluctuations.   Then, we discuss the limits from several observables: the Ly$\alpha$ forest (\S \ref{lya_sec}), the high-redshift galaxy luminosity function (\S \ref{glf_sec}), measurements that constrain early universe star formation from the electron scattering optical depth through reionization (\S \ref{sfr_sec}), and finally from future 21cm observations and the potential shock heating of cosmic gas (\S \ref{cmb_sec}).  While some of these observables are inherently very astrophysical and hence the constraints dependent on modeling, we show that isocurvature fluctuations can result in qualitatively different trends.  Our numerical calculations take
$\Omega_m = 0.308, \Omega_\Lambda = 0.692, \Omega_b = 0.0484, h = 0.678,
\sigma_8 = 0.815$, and $n_s = 0.968$, consistent with the results of \citep{2016A&A...594A..13P}.  When convenient, our calculations will use natural units where $c=\hbar= k_b = 1$.  Cosmological distances and wavenumbers are given in comoving units.   All mass function calculations use the mass function of \citet{ShethTormen2002}.  Even though we are considering non-standard cosmologies, the well-tested universality of the mass function means that \citet{ShethTormen2002} still holds at the 10\% fractional level \citep[and some of us have also have been involved in running simulations testing this]{2009arXiv0908.2702B}.

\section{Isocurvature power from post-inflation axions}
\label{sec:iso}

After perturbative effects break the degeneracy between different $\theta$-vacua, the vacuum misalignment of the ALP translates into a component that behaves like non-relativistic matter with local density \citep{Weinberg1978,Wilczek1978,1990eaun.book.....K}
\begin{equation}
\rho_a(T,\theta_{\rm ini}) = \frac{1}{2} f_a^2  m_a(T) m_a(T_{\rm osc}) \theta^2_{\rm ini} \left(\frac{a(T_{\rm osc})}{a(T)}\right)^3,
\label{eqn:rho}
\end{equation}
where $\theta_{\rm ini}$ is the initial vacuum misalignment angle after symmetry breaking, $a(T)$ is the scale factor, and $m_a(T)$ the axion mass. This formula holds after the axion starts oscillating at an oscillation temperature that we define as $m_a = 3H(T_{\rm osc})$.  Eqn.~(\ref{eqn:rho}) allows for the possibility that the axion temperature is also evolving at $T_{\rm osc}$ as could occur in strongly interacting sectors (as discussed later).  We average $\rho_a(T,\theta_{\rm ini})$ over space, noting that we use the simple relation for the spatial average $\langle \theta_{\rm ini}^2 \rangle = \pi^2/3$, to calculate the average dark matter abundance.  The axion decay constant $f_A$ (which we also refer to as the ``symmetry breaking scale'') will be adjusted to match the observed dark matter abundance.   

Because different causal horizons have different $\theta_{\rm ini}$, this translates into a white spectrum of isocurvature fluctuations in the matter overdensity at times after the field behaves like non-relativistic matter but well into the radiation era with growing mode dimensionless power spectrum of (e.g. \cite{Feix:2019lpo})
\begin{equation}
    \Delta_{\cal S}^2(k) \equiv \frac{k^3}{2\pi^2} P_{\cal S}(k) = A_{\rm osc} \left(\frac{k}{k_{\rm osc}}\right)^3 ~~~\text{at $k < k_{\rm osc}$} ,
\end{equation}
where $P_{\cal S}(k)\equiv  V^{-1} | \tilde \delta_\mathbf{k}|^2$, $V$ is the volume, and $\tilde \delta_k$ is the Fourier transform of the configuration-space dark matter matter overdensity $\delta(\bfx)$ (which we assume to be entirely composed of ALPs such that $\delta(\bfx) = \rho_a(\bfx)/\langle \rho_a \rangle - 1 $), $k_{\rm osc} = a\,H\left.\right|_{T_{\rm osc}}$ is the size of the Horizon when the APL starts to oscillate in its potential \cite{1990eaun.book.....K}, and $A_{\rm osc}$ sets the normalization for which the order-unity fluctuations on the oscillations scale mean $A_{\rm osc} \sim1$.  While irrelevant for this study, at scales $k \gtrsim k_{\rm osc}$ a sharp cut-off is expected as the vacuum misalignment fluctuations have been smoothed out by the Kibble Mechanism \citep{KIBBLE}.  Typical values of $k_{\rm osc}$ are between $100$ and $1000$ $\mathrm{Mpc^{-1}}$ for the ALP masses of $10^{-19}$ and $10^{-17}$ $\mathrm{eV}$, respectively.  The signatures we study are sourced by structures that are coming from an order of magnitude smaller wavenumbers.

Simulations of the QCD axion find that values of the isocurvature variance at initial conditions are $A_{\rm osc} \sim 0.01 - 0.3$ \cite{Feix:2019lpo, Vaquero2019}, somewhat smaller than unity because some of the misalignment power is not in the zero mode and because this signal is diluted by relativistic axions radiated by axionic strings.  However, for our ALP we expect the details that shape $A_{\rm osc}$ to depend on the specific model. 
When we connect our results to the axion mass $m_a$, we take as a fiducial value $A_{\rm osc} = 0.1$, but our results are easily re-scaled to other values.

\begin{figure}
\includegraphics[width=9cm]{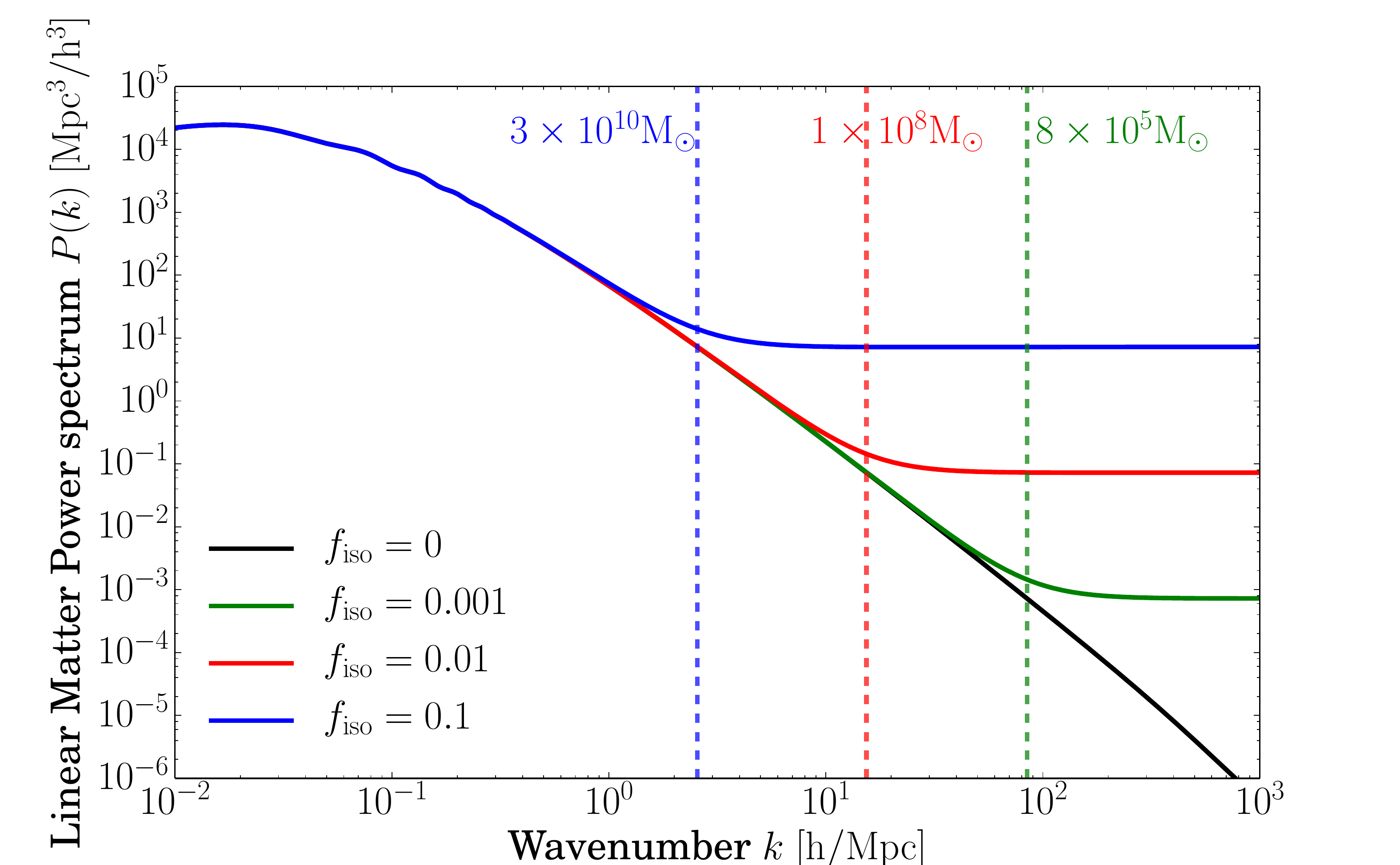}
\caption{The linear matter power spectrum at $z=0$ for different values of $f_{\rm iso}$. The scales where isocurvature contribution become important, relative to the adiabatic power spectrum, are marked with vertical dashed lines and labeled by their $M = 4\pi/3 \rho_m(z=0) k_c^{-3}$, where $k_c$ is the wavenumber where adiabatic and isocurvature fluctuations are equal.
}
\label{fig_5}
\end{figure}

We use the standard growth and transfer function parameterization to model the subsequent evolution of the isocurvature fluctuations (as well as the standard inflationary adiabatic fluctuations).
We parameterize the ioscurvature fluctuations as
\begin{equation}
\Delta^2_{\rm iso}(k,z) = D_{\rm iso}^2(z) T_{\rm iso}^2(k, z) A_{\rm iso} \left(\frac{k}{k_\star}\right)^{3},    
\end{equation}
where $D_{\rm iso}(z)$ is the growth function that tends to a constant deep in the radiation era, and $T_{\rm iso}^2(k, z)$ is the transfer function that is normalized to unity at high-$k$ \footnote{That the isocurvature transfer function limits to unity at high-$k$ is true for the dark matter/ALP transfer function. Whereas the total matter transfer function will be lower due to the effects of Jeans smoothing on the baryons.}. This transfer function is approximately constant for modes that enter the horizon during radiation domination. We take $k_\star = 0.05\;\mathrm{Mpc^{-1}}$ for the pivot scale. Similarly, for the adiabatic fluctuations from inflation  
\begin{equation}
\Delta^2_{\rm ad}(k,z) = D_{\rm ad}^2(z) T_{\rm ad}^2(k,z) A_s \left(\frac{k}{k_\star}\right)^{n_s-1},
\end{equation}
with analogous definitions as for $\Delta^2_{\rm iso}$ except that the adiabatic transfer function is normalized to unity at low $k$.  For our chosen value of $\sigma_8$, $A_s = 2.054 \;\times\;10^{-9}$.  The total matter power at redshift $z$ is the sum of the isocurvature and adiabatic contributions, $\Delta^2_{\rm iso}+\Delta^2_{\rm ad}$. The transfer functions at late times were calculated using {\tt CAMB} Boltzmann code solver \citep{Lewis2002}. Following convention, we define $f_{\rm iso}$ to be the ratio of isocurvature to adiabatic fluctuations at $k_\star = 0.05\;\mathrm{Mpc^{-1}}$:
\begin{equation}
    f_{\rm iso}^2 = \frac{A_{\rm iso}}{A_s}  = \frac{A_{\rm osc}}{A_s} \left(\frac{k_\star}{k_{\rm osc}}\right)^{3},
    \label{eqn:fiso}
\end{equation}
where the second equation uses that deep into the radiation dominated universe $A_{\rm iso}/ {k_\star}^3 = A_{s}/k_{\rm osc}^3$ since $D_{\rm iso} T_{\rm iso} \rightarrow 1$.  In the late time matter power, the ratio of isocurvature-sourced to adiabatic-sourced fluctuations is highly scale dependent, scaling approximately as $k^3$ at high wavenumbers. This is illustrated in Fig.~\ref{fig_5}, where different colours represent different values of $f_{\rm iso}$, with highest value of $f_{\rm iso}$ resulting in highest small scale power. Dashed vertical lines show the mass scale at which the adiabatic and isocurvature contributions to the power spectrum are equal. The contribution of isocurvature fluctuations becomes important at different mass scales, following the approximate scaling of $f_{\rm iso}$ with mass as $M^{1/2}$. This is a direct consequence of the definition of $f_{\rm iso}$ which is fixed on large scales ($k_\star$), and leads to a natural expectation that observables probing smaller mass scales will result in tighter constraints on $f_{\rm iso}$.

We also specify the level of isocurvature by its early time nonlinear scale $k_{\rm nl}$, where $k_{\rm nl} \equiv k_{\rm  osc} A_{\rm osc}^{-1/3}$ such that deep into the radiation era $\Delta^2_{\rm iso} = (k/k_{\rm nl})^3$.  The nonlinear scale represents a more straightforward quantification of the white noise power because it does not convolve in the well-understood amplitude of adiabatic fluctuations and because it does not single out a specific $k_\star$.

One likely scenarios is that the $m_a$ does not exhibit strong temperature dependence in the early Universe.  This limit applies to ALPs whose mass is acquired by nonperturbative effects associated with the perturbative gauge couplings in GUT theories \citep{2017PhRvD..95d3541H}. In this case, the non-perturbative mass is exponentially suppressed relative to the symmetry breaking scale and the ALP field obtains its zero-temperature mass at $T\gg T_{\rm osc}$.  We also consider a QCD-like case of a asymptotically-free strongly interacting sector where the non-perturbative effects increase with decreasing temperature until the temperature reaches the confinement scale, $\Lambda$; evolution of the mass occurs even after the ALP behaves like non-relativistic matter if $\Lambda < T_{\rm osc}$, with the final mass equal to $m_A = \Lambda^2/f_a$. The ALP mass evolution can be characterized at $T \lesssim T_{\rm osc}$ by
\begin{eqnarray}
    m_a(T) &=&  m_a \left(\frac{\Lambda}{T}\right)^n ~~~~\text{for $T>\Lambda$},\\
     m_a(T) &=&  m_a ~~~~\text{otherwise},
\end{eqnarray}
where we use the notation that $m_a$ without an argument is the zero temeprature mass and where $n$ parameterizes the temperature dependence of the instanton effects.  The case $n=4$ mimics the scaling found for the QCD axion, but the details of this scaling will depend on the strong sector.  For $n=0$ perturbative case, we note that this parameterization still holds (trivially). 

With this parameterization,
\begin{eqnarray}
  T_{\rm osc} &=& 3 \left(
 \frac{10}{\pi^2 g_{\rm eff}}\right)^{1/4} [m_A(T_{\rm osc}) M_P]^{1/2} \\
 &\propto& \langle \theta_{\rm ini}^2 \rangle^{-\frac{n}{8+3n}} m_a^{\frac{4+n}{8+3n}}, \\
    k_{\rm osc} &=& a_{\rm osc}  H(T_{\rm osc}) = \frac{T_{\rm cmb,0}}{T_{\rm osc}}\frac{ m_A(T_{\rm osc})  }{3} \propto T_{\rm osc} ,\\
        f_A &\propto& \langle \theta_{\rm ini}^2 \rangle^{-2/(8+3n)} m_a^{-(2+n)/(8+3n)} 
    \label{eqn:kosc},
\end{eqnarray}
where $M_P = 1/\sqrt{8\pi G}$ is the reduced Planck mass, and $ T_{\rm osc}$ evaluates to $1-100$keV for $m_A$ of interest, indicating $g_{\rm eff}\approx 3.4$.  For the proportionality relations, we have eliminated the $\Lambda$ dependence in favor of $m_a$ and $f_A$.
We note that at fixed $m_A(T_{\rm osc})$ the amplitude of isocurvature fluctuations does not depend on $n$, and our constraints in Fig.~\ref{fig_constraints} translate to $m_A(T_{\rm osc}) =  10^{-20} -10^{-17}$eV.  For our $n=4,10$ models in Fig.~\ref{fig_constraints}, the particle mass increases by $3,4$ orders of magnitude to reach $m_a$ at $T= \Lambda$.

Fig.~\ref{fig_constraints} foreshadows the constraints we find in the following sections  in the $m_a-f_{\rm iso}$ plane.  The different horizontal limits show the upper limit on $f_{\rm iso}$, bounding the viable parameter space to be below the curves. The $n=0$ corresponds to the most likely case where the mass is established well before the particle commences oscillations, and the QCD axion yields a scaling with $n= 4$. The dots on the lines correspond to the values of the decay constant $f_A$ for those models (colour coded to match the lines), while the shaded regions around the lines correspond to the uncertainty in the value of $A_{\rm osc}$. The sold lines themselves were evaluated at the value of $A_{\rm osc} = 0.1$.

\begin{figure}
\includegraphics[width=9cm]{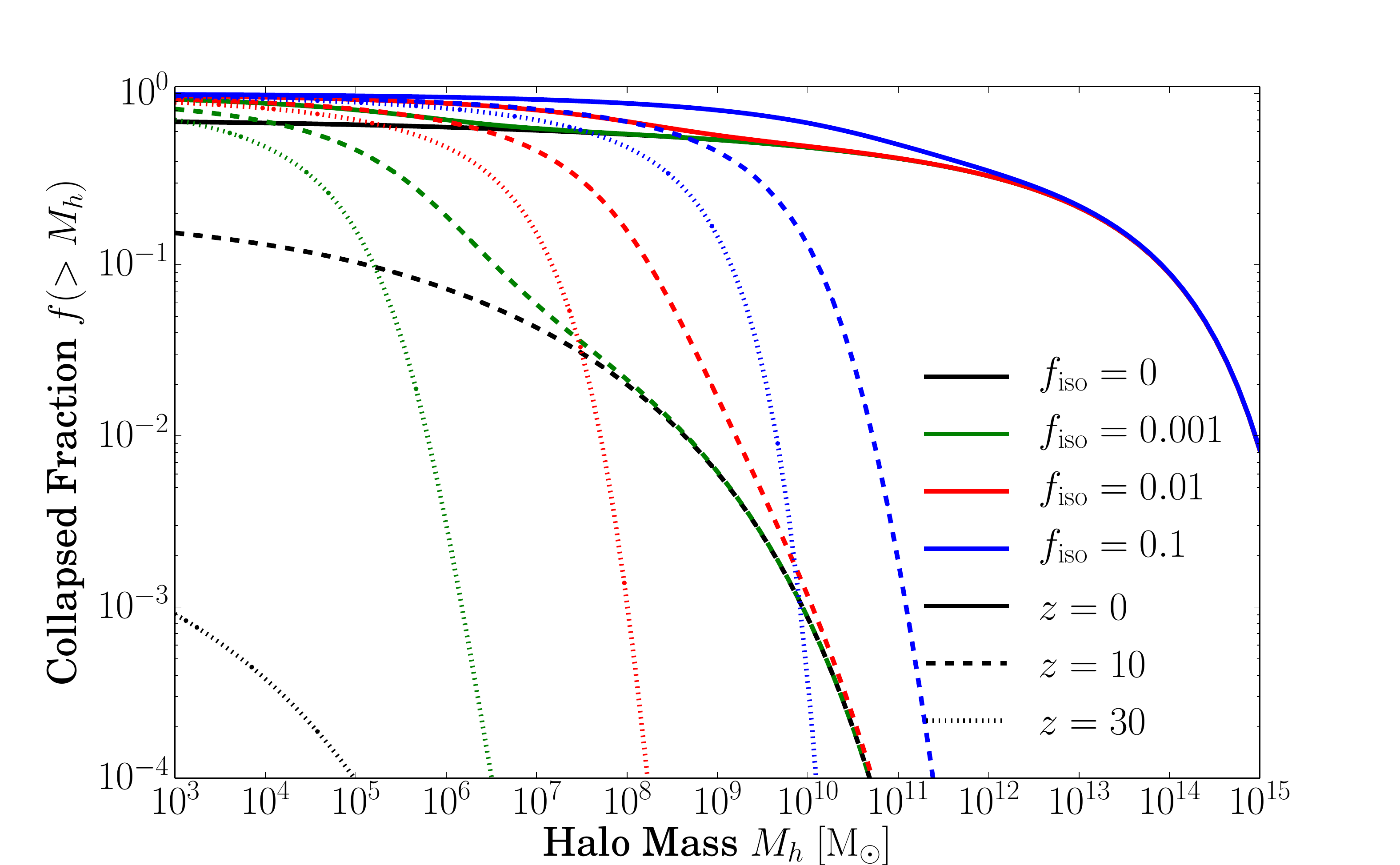}
\caption{The collapsed fraction in halos above a mass of $M_h$.  Different colours show the collapsed fraction for isocurvature fractions of $f_{\rm iso}=0.1$ (blue), $f_{\rm iso}=0.01$ (red), $f_{\rm iso}=0.001$ (green) and $f_{\rm iso}=0$ (black). The line styles differentiate redshifts.  
}
\label{fig_7}
\end{figure}

\section{Lyman-$\alpha$ forest}
\label{lya_sec}

The Lyman-$\alpha$ forest is used to infer the initial conditions using significantly smaller comoving scales than other large-scale structure observables, to 3D wavenumbers of $k\approx 10-100$Mpc$^{-1}$ \citep{2009RvMP...81.1405M, mcquinn-review}.  The Lyman-$\alpha$ forest circumvents many of the difficulties of modeling structure formation at these nonlinear scales by being sensitive exclusively to low-densities ($\Delta \sim 1$ as the absorption of higher densities is saturated; \cite{2018JCAP...04..026I}) where our nonlinear models for the cosmic web appear to be under control \citep{1994ApJ...437L...9C, 1996ApJ...471..582M, 1996ApJ...457L..51H} and where astrophysical processes appear to be less of a contaminant \citep[e.g.][]{mcquinn-review}.  Indeed, the forest has been used to place the tightest constraints on  small-scale cutoff in the spectrum of primordial matter fluctuations, which may owe to the free streaming of warm dark matter and the de-Broglie wavelength of fuzzy dark matter \citep{2006PhRvL..97s1303S, 2005PhRvD..71f3534V,  2017PhRvL.119c1302I}. In the context of ALPs, combining the Ly$\alpha$ constraints with the limits on the isocurvature fluctuations from the CMB can lead to interesting bounds on the tensor-to-scalar ratio \citep{Kobayashi2017}.

A typical Ly$\alpha$ forest analysis is sensitive to 1D wavenumbers between 0.1 and 10 $\mathrm{Mpc/h}$, which would naively lead to a typical mass of $10^8\;\mathrm{M_\odot}$ (see Fig.~\ref{fig_5}). However, The non-linear mapping from the 3D density field to the 1D flux field in the quasar spectra makes the Ly$\alpha$ forest sensitive to even smaller wavenumbers (see e.g. \citep{murgia19}). Additionally, the non-linearity of the gravitational evolution does not dominate over the clustering signal at high redshifts, which helps to better constrain cosmology at a given scale.

The forest is also sensitive to an enhancement in power as would occur from the white isocurvature fluctuations from axions in the post-inflation scenario.  Indeed, the allowed level of enhancement has been constrained in the context of primordial black holes, which also may have a white spectrum \cite{2003ApJ...594L..71A, murgia19}. Conveniently, the adiabatic plus white-noise simulations run for the primordial black holes in \cite{murgia19} are the same as would be run in the context of ALP isocurvature perturbations, the difference comes in the interpretation of the isocurvature amplitude and how it is linked to the actual physical model. In particular, Murgia and coworkers \cite{murgia19} find that the isocurvature fraction of $f_{\rm iso} = \sqrt{A_{\rm iso}/A_s}$ at the pivot scale of $k = 0.05\;\mathrm{Mpc^{-1}}$, should be lower than $0.004$ at $2\sigma$ confidence level when adopting conservative priors on the thermal history. This constraint can be remapped to our models by solving Eqs.~\ref{eqn:fiso} and~\ref{eqn:kosc} for a given ALP mass evolution model. The relation between $f_A$ and $m_a$ is fixed by assuming that all of dark matter is composed of the axion-like particle. This gives a lower bound on the mass of the ALP of $m_A >2\times 10^{-17}\;\mathrm{eV}$ for the most natural case of no mass evolution after the axion starts oscillating ($n=0$).
This constraint further shows that the forest is effectively able to probe structure in the dark matter to mass scales as small as $\sim 3\times10^7\;\mathrm{M_\odot}$ (using Fig.~\ref{fig_5}); a number that is helpful for putting the forest in context with the other constraints we discuss.

Figure~\ref{fig_4} shows the constraints from the forest. A primary result of this paper is that we find the Ly$\alpha$ forest is more constraining than other probes, although future observations of the high-redshift universe using redshift 21cm radiation may ultimately be more constraining.

\section{Galaxy luminosity function} 
\label{glf_sec}

Small galaxies are a second observable that has been used to constrain the primordial fluctuations on small scales, with observations both probing them as satellite galaxies to the Milky Way \citep{2017ARA&A..55..343B} and at high redshifts when they are forming the bulk of their stars \citep{2001ApJ...558..482B, 2013MNRAS.435L..53P}. Since the white-noise isocurvature fluctuations in our ultra-light axion models dramatically increase fluctuations on small scales, such scenarios may predict a large increase in the number of low-luminosity galaxies.  Foreshadowing the result of this section: For galaxies that are directly observable in the future, we find that this enhancement is small for the $f_{\rm iso}$ allowed by the forest, although in \S~\ref{sfr_sec} we show that for smaller galaxies (whose effects can only be indirectly probed via their ionization and enrichment) the enhancement can be more substantial. 

To model the enhanced number of small galaxies, we use a simple but successful model for star formation where the predicted number density of galaxies $n_g$ per UV luminosity between $L$ and $L+dL$ is related to the halo mass function $dn_h/dM_h$ by 
\begin{equation}
\phi(L)\equiv\frac{dn_g}{dL} = \frac{dn_h}{dM_h} \frac{dM_h}{dL},
\label{eq:phiAB}
\end{equation}
This model assumes the common one-to-one mapping between halo mass and observed UV luminosity described by ${dM_h}/{dL}$.  As this function has significant astrophysical uncertainty, we will use  qualitatively different shapes for the galaxy luminosity function, ${dn_g}/{dL}$, as a signature that a given axion cosmology is excluded.

To calculate the terms in eqn.~\ref{eq:phiAB}, we use the Sheth-Tormen mass function \citep{ShethTormen2002} to model ${dn_h}/{dM_h}$ \footnote{We have checked that the results are not sensitive to the choice of the mass function by also investigating a mass function specifically calibrated to simulations at high redshift \citep{Trac2015}.}  The `universality' of the halo mass function makes it likely that the same mass function should be a good approximation to cases with isocurvature fluctuations \citep[e.g.]{2007ApJ...671.1160L, 2009arXiv0908.2702B}. Additionally, we adopt a common assumption that a galaxy's star formation rate is proportional to its gas accretion rate, $\dot{M}_b$, with proportionality constant $f_\star(M_h, z)$ called stellar efficiency.  Note that the star formation rate directly maps to the UV luminosity of the galaxy.  We follow \citet{furlanetto16} to calculate $f_\star$, who calculate it from an analytic model that considers energy regulated stellar feedback process plus virial shocking. In this model, the stellar efficiency of the baryons peaks at around $M_h = 10^{11.5}\;\mathrm{M_\odot}$, where it reaches the values of just below $0.05$. This efficiency has a steep tail towards smaller masses, reaching $10^{-3}$ by $M_h = 10^{8}\;\mathrm{M_\odot}$.  One worry, which we will address, is that this efficiency depends on uncertain astrophysics and so any differences we find may not be distinguishable.

To model the gas accretion rate $\dot{M}_b$, numerical results are typically obtained from cosmological simulations (e.g. \cite{mcbride09}), but for the isocurvature case, $\dot{M}_b$ has not been determined using simulations. However, the time evolution of the halo accretion rate is driven largely by the time evolution of the mass variance $\sigma(M)$ (see e.g. \citep{Correa2015}). We set
\begin{equation}
    \frac{d\ln{M_b}}{dt} = \left| \frac{d\ln{\sigma}}{d\ln{M_h}} \right|^{-1} \frac{d\ln{D}}{dt},
\end{equation}
and $D$ is the growth function. This allows us to build a consistent approach to calculating the gas accretion for any $f_{\rm iso}$. Our results on the gas accretion are in good agreement \cite{Correa2015} in the limit they consider of $f_{\rm iso}=0$.

\begin{figure}
\includegraphics[width=9cm]{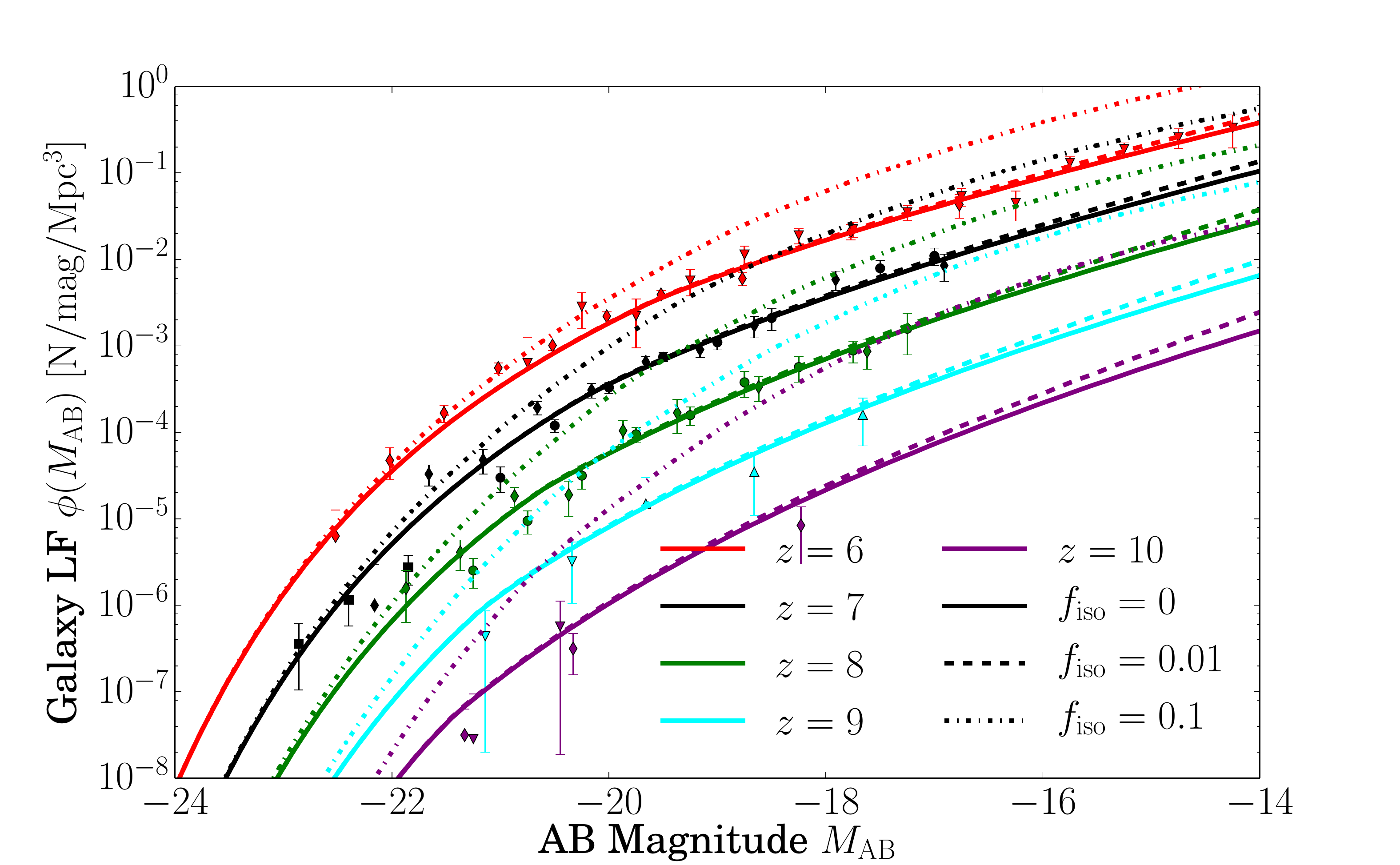}
\caption{The effect of white-noise isocurvature fluctuations on the galaxy luminosity function. The various linestyles shows different levels of isocurvature fluctuations, with colour indicating redshift. Overplotted is a compilation of observational data ranging over typical redshifts probed by the future surveys. We consider $f_{\rm iso} =0.1$ to be ruled out by these observations as by this value the luminosity function has a qualitatively different behavior, especially at the highest redshifts probed.}
\label{fig_1}
\end{figure}

Fig.~\ref{fig_1} shows the resulting comparison of the galaxy luminosity function. Our model is compared to the measurements of \citep{McLure2013,Bowler2017, Bouwens2015, Bouwens2016}, but also include the $z=6$ lensed galaxy sample of \citep{Bouwens2017} that extend the measurement to fainter immensities.  We use the standard convention of writing the UV luminosity in terms of absolute AB magnitude where $M_{\rm AB} = -2.5 \log_{10}(L_{\rm UV}) +M_{\rm ref}$ where $M_{\rm ref}$ is a constant. We have not performed any dust correction at this stage, as the typical corrections \citep[e.g.][]{Smit2012} are only significant for the higher mass systems, and leads to a shallower relation between the halo mass and the UV magnitude \footnote{This effect may weaken our constraints from the galaxy luminosity function if lower mass galaxies are substantially dust absorbed.}.

However, including isocurvature fluctuations, even at level already excluded by Ly$\alpha$ forest of $f_{\rm iso} = 0.01$, only results in a small signal at a lower end of the luminosity function. This is mainly due to the fact that even the observed high-redshift galaxies behind cluster lenses reside in $>10^9M_\odot$ halos in our models.  In contrast, the Ly$\alpha$ forest is sensitive to scales of $M\sim 10^8\,M_\odot$, as illustrated in Fig.~\ref{fig_5}. We find that current observations of the high-redshift luminosity function rule out $f_{\rm iso} > 0.1$, as this leads to a large qualitative change that likely cannot be mimicked by the large astrophysical uncertainty in our star formation efficiency model.  One can already start to see this large effect for the $f_{\rm iso} =0.05$ model in Fig.~\ref{fig_5}. These limits translate into a lower bound on the ALP mass to be $m_a > 10^{-19}\;\mathrm{eV}$.

Future observations at higher redshifts would help in discriminating between different isocurvature models, and could potentially provide constraints comparable to the ones derived from the small scale structure of the Ly$\alpha$ forest. Namely, the James Webb Space Telescope (JWST) is able to go a few magnitudes deeper at $z\approx 6$, and more importantly has the infrared sensitivity that allows better constraints at higher redshifts. With lensed galaxy samples, JWST should be able to place similar constraints to HST at $z=6$ (reaching to absolute magnitudes of $M_{\rm AB} = -14$) but all the way to $z=10$, constraining $f_{\rm iso} \sim 0.01$.  Unfortunately, astrophysical uncertainties require a qualitative change in behavior, making it difficult to probe beyond $f_{\rm iso} = 0.01$.  Thus, the Ly$\alpha$ forest is likely to always provide a more sensitive probe than direct measurements of galaxy luminosity functions. 

\section{High-redshfit star formation rate and reionization}
\label{sfr_sec}

Though we find that the galaxy luminosity function is not competitive with the Ly$\alpha$ forest, the collapsed fraction of halos that can form stars can be orders of magnitude larger than the $f_{\rm iso}=0$ prediction at $z=10$, and this difference is even larger at higher redshifts, if we take $f_{\rm iso} = 0.01$ -- comparable to the constraint coming from Ly$\alpha$.  This is illustrated in Fig~\ref{fig_7}, noting that stars can only form in halos with $M_h \gtrsim 10^{7-8}M_\odot$ if the gas condenses by cooling via atomic transitions and $M_h \gtrsim 10^{5-6}M_\odot$ halos if instead by molecular ones.  Unfortunately, the direct luminosity function measurements with HST (and in the future with JWST) are not sufficiently sensitive to detect the stars/galaxies that likely lie in these diminutive halos.  However, the enhanced extremely high-redshift star formation from an increased abundance of these small halos could also heat and ionize the cosmic gas (and their UV photons can pump the 21cm line) in a manner that may allow constraints on $f_{\rm iso}$.  There is also some indirect evidence that the smallest galaxies contribute disproportionately to the ionizing photons that escape into and hence ionize the IGM  \citep[e.g.]{Haardt2012}, which would make our mass-independent escape in what follows conservative.

To illustrate just how much isocurvature fluctuations could change the mass in halos that are massive enough to host stars, we calculate the fraction of mass that is collapsed in halos with masses above $M_h$ using Extended Press-Schechter theory \citep{Press1974,Bond1991}.  This yields
$f_{\rm coll}(>M_h) = \mathrm{erfc}\left( {\nu(M_h)}/{\sqrt{2}} \right)$,
where $\mathrm{erfc}(x) \equiv \pi^{-1/2} \int_x^\infty dx \exp[-x^2]$ and $\nu \equiv \delta_c/\sigma(M, z)$ and $\sigma(M, z)$ is the standard deviation of the density in a spherical top-hat Lagrangian volume with mass $M$.  The virial temperature of halo (the characteristic temperature the gas can shock heat) is the property of a halo that sets whether its gas can cool and form stars rather than the halo mass.  The two are related by $M_h \propto \left[a T_{\rm vir}\right]^{3/2}$ -- at higher redshifts the same virial $T_{\rm vir}$ halo has smaller $M_h$. The isocurvature fluctuations with constant power spectrum on small scales during the matter dominated epoch this leads to $\sigma^2(M) \propto a^2 M^{-1}$, whereas for $f_{\rm iso} = 0$ we have $\sigma^2(M) \propto a^2 \log[M]$.  The result is that the redshift evolution of the collapsed fraction at fixed virial radius is \emph{much} flatter for masses where isocurvature fluctuations dominate, with the difference given by
\begin{eqnarray}
f_{\rm coll}(>T_{\rm vir}) &=& {\rm erfc}\left[1.7 ~Z_{10}\right ] \text{~~~~~~~~~~~~~~~for~adiabatic;} \nonumber\\
f_{\rm coll}(>T_{\rm vir}) &=& {\rm erfc}[1.0 ~f_{{\rm i},-2}^{-1} Z_{10}^{1/4} T_{{\rm vir},4}^{3/4} ] \text{~~ ~isocurvature,}\nonumber
\end{eqnarray}
where $f_{{\rm i},-2} \equiv f_{\rm iso}/10^{-2}$, $Z_{10} \equiv (1+z)/10$ and $T_{{\rm vir},4} \equiv  T_{\rm vir}/10^4~{\rm K}$.\footnote{The full dependence on redshift and virial temperature for the adiabatic case is roughly $f_{\rm coll}(>T_{\rm vir}) = {\rm erfc}\left[2.04 \times Z_{10} \left\{\ln\left(4.6 \times Z_{10} T_{{\rm vir},4}^{-1}\right)\right\}^{-1/2}\right ]$, but the logarithmic dependence only adds a small correction to the redshift evolution.}

The former function falls off exponentially with increasing redshift for rare (large $\nu$) objects noting asymptotic form $\sqrt{\pi} \mathrm{erf} (x) = \exp[-x^2](x^{-1} +{\cal O}(x^{-3}))$, whereas the latter (while still exponentially sensitive once the argument becomes greater than unity) is much flatter, allowing halos that can cool at much higher redshifts. 

\begin{figure}
\includegraphics[width=9cm]{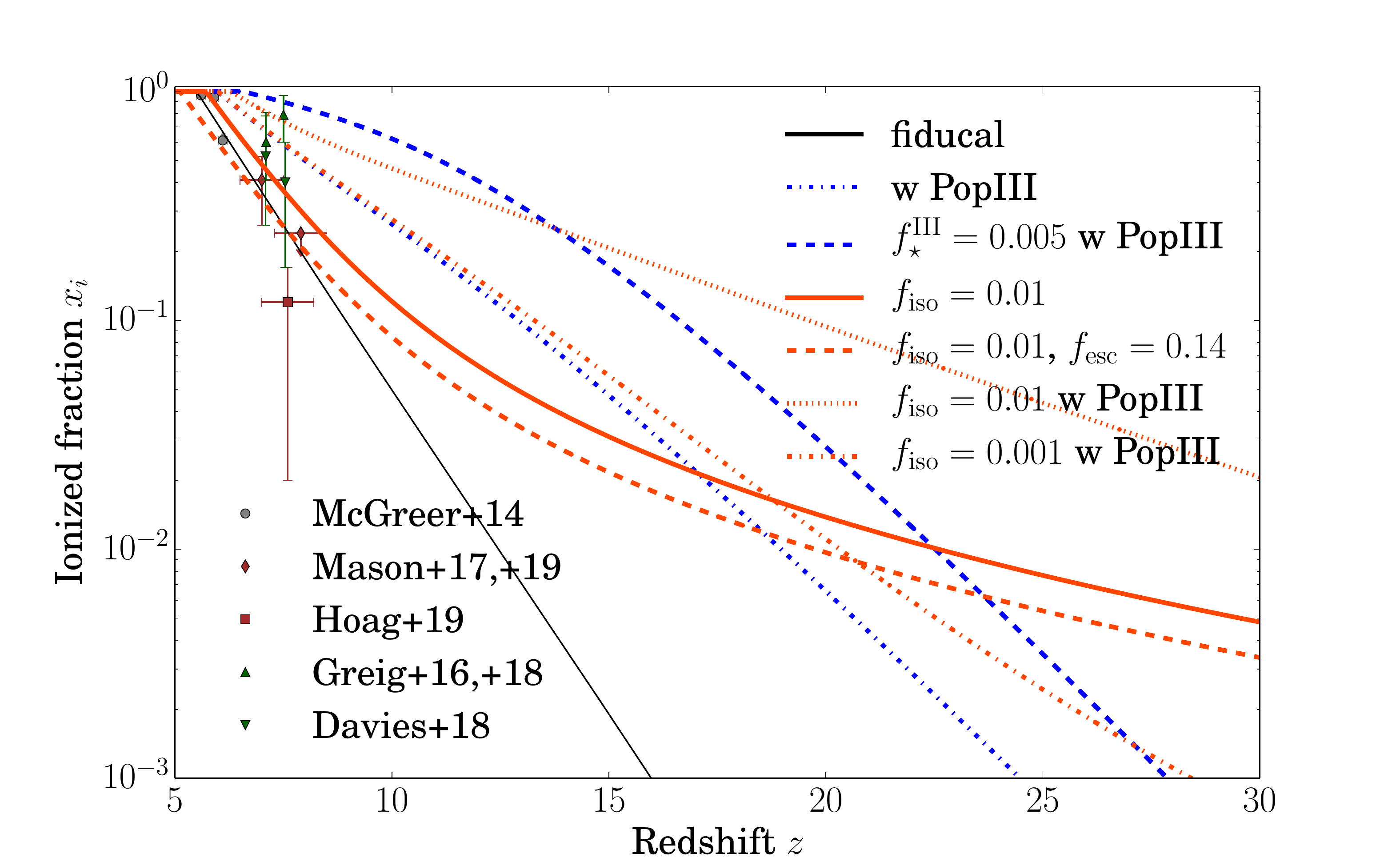}
\caption{The evolution of the ionized fraction of the intergalactic gas during reionization. The fiducial model assumes a given stellar efficiency described in Sec.~\ref{glf_sec}. The effect of axion isocurvature fluctuations is shown for various values of $f_{\rm iso}$ and also varying assumptions about the escape fraction of Pop-II stars (orange dashed) or including a contribution from Pop-III stars (blue and orange dot-dashed). Overplotted is a compilation of observational constraints on the ionized fraction coming from Ly$\alpha$ dark pixels (grey), Ly$\alpha$ emitters (brown) and QSO damping wings (green). }
\label{fig_2}
\end{figure}

An enhancement in the number of star forming halos in the manner of our white isocurvature fluctuations should lead to an enhanced number of hydrogen ionizing photons, causing the reionization of the Universe to start earlier and be a much more prolonged process. Such a reionization history would be constrained by direct estimates of the ionized fraction using quasar spectra and Lyman-$\alpha$ emitters. The ionized state of the intergalactic gas can be measured through the time evolution of the volume-averaged ionized fraction, that depends on the balance between recombination and ionization due to photo-ionization \citep{Sun2016},
\begin{equation}
    \frac{dx_{\rm i}}{dt} = \frac{d\left(\zeta f_{\rm coll}\right)}{dt} - {\bar n}_H(t)\,\alpha_{\rm re}(T_e)\,C_{\rm HII}\,x_{\rm i},
    \label{eq:xion}
\end{equation}
where $\zeta = A_{\rm He} f_\star f_{\rm esc} N_{\gamma}$ is the ionizing efficiency: a product of the correction factor for singly ionized helium, $A_{\rm He} \approx 1.22$; the star formation efficiency, $f_\star$; the escape fraction of ionizing photons, $f_{\rm esc}$; and the average number of ionizing photons produced per stellar baryon, $N_{\gamma}$. In the recombination term, the number density of hydrogen, ${\bar n}_H$, is time dependent as ${\bar n}_H = {\bar n}_H(z=0)\left(1+z\right)^3$ at redshift $z$; the recombination rate, $\alpha_{\rm re}$, is temperature dependent such that $\alpha_{\rm re}(T_e) = 2.6\;\times\;10^{-13}\,\left(T_e/10^4\;\mathrm{K}\right)^{0.76}\,\mathrm{cm^3\,s^{-1}}$, at the electron temperature $T_e$; and the volume-averaged clumping factor is defined to be $C_{\rm HII}\equiv \langle n_e^2\rangle/\langle n_e \rangle^2$. 

A rough approximation during HI reionization \citep{Shull2012,Sun2016} is to fix $C_{\rm HII} = 3$, and $T_e = 10^4\;\mathrm{K}$. It would be naturally to expect a redshift evolution of the clumping factor (see e.g. \citep{Haardt2012}), which might change the reionization history. In our simple scenario, chaning the value of the clumping factor to 5 (1) leads to a largely redshift-independent change in the ionized fraction in our calculations by a factor of 0.8 (1.4) (at least at high redshifts). The value of the mean number of ionizing photons produced, $N_{\gamma}$, depends in the initial mass function and metalicity of the stellar population. We use $N_{\gamma} = 4,000$ for Population II (Pop-II) stars, assuming Salpeter IMF and 5\% of the solar metalicity (although the results are weakly sensitive to these choices at least assuming empirically motivated IMFs).  Pop-II stars are the second generation of stars that are born in metal enriched gas and likely have properties similar to stars observed at low-redshifts. Unless otherwise stated we use the escape fraction of 20\% for the Pop-II stars. In the fiducial Pop-II model we assume all halos above $M_{\rm min}$ form stars, and at each redshift the value of $M_{\rm min}$ is fixed to the mass at the virial temperature of $T_{\rm vir} = 10^4\;\mathrm{K}$.
The basic photo-ionization rate can be evaluated using the halo mass accretion rates discussed in (\S \ref{glf_sec}),
\begin{equation}
    \frac{d\left(\zeta f_{\rm coll}\right)}{dt} = A_{He}\, N_\gamma\, f_{\rm esc}\, \int_{M_{\rm min}}^\infty \frac{dM_h}{{\bar \rho}_m} n(M_h) f_\star {\dot M}_h ,
    \label{eq:zeta_II}
\end{equation}
where $f_\star$ is the mass-dependent stellar efficiency and $n(M_h)$ is the halo mass function.

In the context of the early star formation, a Population III (Pop-III) stellar contribution is often discussed, which is the first generation of stars which are born metal free and expected to be more massive. Since this contribution is at present largely unconstrained \citep{Visbal2015}, we adopt a toy model to characterize their effect on the progression of the reionization. In this case an additional photo-ionization term is added, mimicking the structure of $d\left(\zeta_{\rm III} f_{\rm coll}\right)/dt$, but with the ionizing efficiency characteristic of the Pop-III models.  Namely, following Eqn.~(\ref{eq:zeta_II}) we write down the Pop-III photo-ionization rate as
\begin{equation}
    \frac{d\left(\zeta_{\rm III} f_{\rm coll}\right)}{dt} = A_{He}\, N_{\gamma}^{\rm III}\,\, \int_{M_{\rm min}^{\rm III}}^{M_{\rm min}} \frac{dM_h}{{\bar \rho}_m} n(M_h) f_\star^{\rm III} {\dot M}_h .
    \label{eq:zeta_III}
\end{equation}
The integration is only over halos where molecular cooling is efficient and atomic is not (as atomic leads to our normal mode of star formation), i.e. between $T_{\rm vir}= 500\;\mathrm{K}$ ($M_{\rm min}^{\rm III}$), warm enough to excite rotational transitions of molecular hydrogen, and the mass at the virial temperature of $10^4\;\mathrm{K}$ ($M_{\rm min}$). We use $N_{\gamma}^{\rm III} = 40,000$ as anticipated for the hotter photospheres of these metal free stars\citep{Bromm2001}, and assume that all ionizing photons escape as anticipated for star formation in these diminutive halos.  We also take a stellar efficiency of $f_{\star}^{\rm III} = 5\times 10^{-4}$, although the escape of ionizing photons can be pulled into this parameter.  This efficiency is on the lower end of what is typically used in the literature \citep{Trenti2009,Visbal2018}, with most commonly used values being $10^{-3} - 10^{-2}$. However, in our simplifed model, our fiducial value of $f_{\star}^{\rm III}$ leads to the star formation rate density of Pop-III stars comparable to that of \citep{Visbal2015} (see our endnote \footnote{The star formation rate density in our model peaks at around $2\times 10^{-4}\;\mathrm{M_\odot yr^{-1} Mpc^{-3}}$ at redshift of 15, and falls off towards higher redshifts (e.g. $10^{-6}\;\mathrm{M_\odot yr^{-1} Mpc^{-3}}$ at redshift of 35), behaviour quantitatively very similar to that found in \citep{Visbal2015}. This is true despite different star formation efficiency assumed in our model compared to \citep{Visbal2015}, because the minimum mass in which molecular cooling can lead to Pop-III star formation is lower in our model, compared to the that of \citep{Visbal2015}. In \citep{Visbal2015} the numeric value of the minimum mass is obtained from CDM simulations and corresponds to roughly $T_{\rm vir} = 1000$ K. See Eqn.~\ref{eq:M_H2} as the minimum does not just set the absolute minimum but also what halos are affected by the Lyman-Werner background.}).

Once enough stars form in the Universe, the $\sim 11$eV Lyman-Werner radiation they produce dissociates molecular hydrogen, turning off cooling in molecular cooling halos and preventing the formation of further Pop-III stars \citep{Haiman1997, Haiman2000}. To model this we follow \citep{mcquinn12,Visbal2015,Mebane2018}, where we modify the lower integration limit ($M_{\rm min}^{\rm III}$) in Eqn.~(\ref{eq:zeta_III}) to also include self-regulations due to Lyman-Werner background. The numerical calculations of \citep{Machacek2001, Wise2007} found that the gas is able to cool in halos with mass
\begin{equation}
    M_{\rm min}^{\rm III} = M_h\left(T_{\rm vir}=500\;\mathrm{K}\right) \left[ 1+ 6.69\,F_{LW,21}^{0.47} \right] ,
    \label{eq:M_H2}
\end{equation}
where $F_{LW,21}$ is the Lyman-Werner intensity integrated over solid angle in units of $10^{-21}\;\mathrm{erg\, s^{-1}\,Hz^{-1}\,cm^{-2}}$. To estimate the Lyman-Werner intensity given a star formation rate (${\dot \rho}_{\rm SFR}$), we use the relations of \cite{Visbal2015,Mebane2018}
\begin{equation}
    F_{LW,21} = 7.22\, \frac{(1+z)^3}{H(z)}\, e^{-\tau_{LW}} \left(N_{\rm LW}^{\rm II}{\dot \rho}_{\rm SFR}^{\rm II} + N_{\rm LW}^{\rm III}{\dot \rho}_{\rm SFR}^{\rm III} \right) ,
    \label{eq:F_LW}
\end{equation}
where $H(z)$ is the Hubble rate of expansion, and $\tau_{LW}$ is the intergalactic opacity for the Lyman-Werner photons which can be $1 - 2$ in the absence of dissociations \citep{ricotti2001} and can be larger once the first HII regions have formed \citep{Johnson2007}. We use $\exp\left(-\tau_{LW}\right) = 0.5$, however we note that in the isocurvature model the value of $\tau_{LW}$ might increase due to more small scale structure obscuring the Lyman-Werner background. 

The number of Lyman-Werner photons produced per baryon in stars is taken to be $N_{\rm LW}^{\rm II} = 9,690$ for Pop-II stars, and $N_{\rm LW}^{\rm III} = 100,000$ for Pop-III stars \citep{Mebane2018}. The value of ${\dot \rho}_{\rm SFR}$ is modelled through Eqns.~(\ref{eq:zeta_II}) and~(\ref{eq:zeta_III}), such that ${\dot \rho}_{\rm SFR} = f_\star d\left(f_{\rm coll}\right)/dt$. We use an iterative process to determine the value of ${\dot \rho}_{\rm SFR}^{\rm III}$ that satisfies Eqns.~(\ref{eq:zeta_III}), (\ref{eq:M_H2}) and (\ref{eq:F_LW}).

We also multiply Eqn.~(\ref{eq:zeta_III}) by $(1-x_i)$ to account for the photo-heating. This term only becomes important towards the end of reionization at lower redshifts, but prevents the Pop-III photo-ionization term from resulting in overly large optical depth contribution in the range of $10-15$.
The functional form of the above model is an approximate way to characterize the self-regulation of the Pop-III stellar population in the early Universe. Simpler models regulated by the average ionized fraction (e.g. \citep{Miranda2017}) give very similar results. We would also comment that relations in \citep{Visbal2015,Mebane2018} that we use to derive Eqns.~(\ref{eq:M_H2}) and (\ref{eq:F_LW}) were empirically determined from CDM simulations. An approach based on simulations is most likely required to model the details of the Pop-III star formation history in the presence of the isocurvature fluctuations.

However not including any self-regularization leads to larger ionized fractions earlier in its evolution, which violate the observational constraints shown in Fig.~\ref{fig_2}, as well as the integrated optical depth from Planck (see below). Thus some form of self-regularization is important to implement, but the exact details of the model do not change the quantitative picture that including the isocurvature fluctuations leads to a slower decrease of the ionized fraction at higher redshifts, compared to just Pop-III star formation, which is illustrated in Fig.~\ref{fig_2}.

Fig.~\ref{fig_2} shows how the ionized fraction evolves in the redshift range probed by the measurements. Current observations from a variety of sources are plotted on Fig.~\ref{fig_2}: Ly$\alpha$ dark pixels (\citep{McGreer2015} in grey), Ly$\alpha$ emitters (\citep{Mason2018,Hoag2019,Mason2019} in brown), and QSO damping wings (\citep{Greig2017,Greig2019,Davies2018} in green). The fiducial model (black solid line), uses only Pop-II photo-ionization rates, with $f_{\rm esc} = 0.2$ and no isocurvature fluctuations ($f_{\rm iso}=0$). The effect of including axion isocurvature fluctuations (red lines) exhibits a distinctly longer tail of reionization, where the ionized fraction starts to increase much earlier and at a steadier rate than for the no iosocurvature case. At lower redshift, where the ionized fraction can be currently estimated, the effect of the isocurvature fluctuations is slightly degenerate with the escape fraction of Pop-II stars (red dashed line). 

On the other hand, the effect of Pop-III stars is prominent at higher redshifts (green dot-dashed line), and in tandem with the isocurvature fluctuations (dot-dashed red line) can create a boost to the ionized fraction such that it evolves much slower between redshifts of $25$ and $10$, potentially creating a strong observable signal of the isocurvature modes in the future observations. However, enhancing the star formation efficiency for Pop-III stars to $0.005$ as used in \citep{Visbal2018} increases the ionized fraction evolution even without isocurvature fluctuations (green dashed line in Fig.~\ref{fig_2}), making it not obvious that the astrophysics of star formation can be robustly disentangled from $f_{\rm iso}$. Nevertheless, at high enough redshift all our isocurvatore models cross the green-dashed line in Fig.~\ref{fig_2} that corresponds to this extreme case of Pop-III stellar efficiency. This is the unique signal of the isocurvatore models in the ionization history, resulting from the nearly redshift-independent collapse fraction in such models.

Future observations by ground based surveys (e.g. UKIDSS \citep{Lawrence2007}; VIKING \citep{Edge2013}; VHS \citep{McMahon2013}; UHS \citep{Dye2018}) and wide-field surveys (e.g. Euclid, WFIRST, WEAVE, J-PAS) in combination with high signal-to-noise spectra from JWST would be more sensitive to the differences between the models. In particular measuring the ionized fraction during the cosmic dawn epoch ($z>15$) can lead to stronger constraints on the isocurvature fluctuations.

\begin{figure}
\includegraphics[width=9cm]{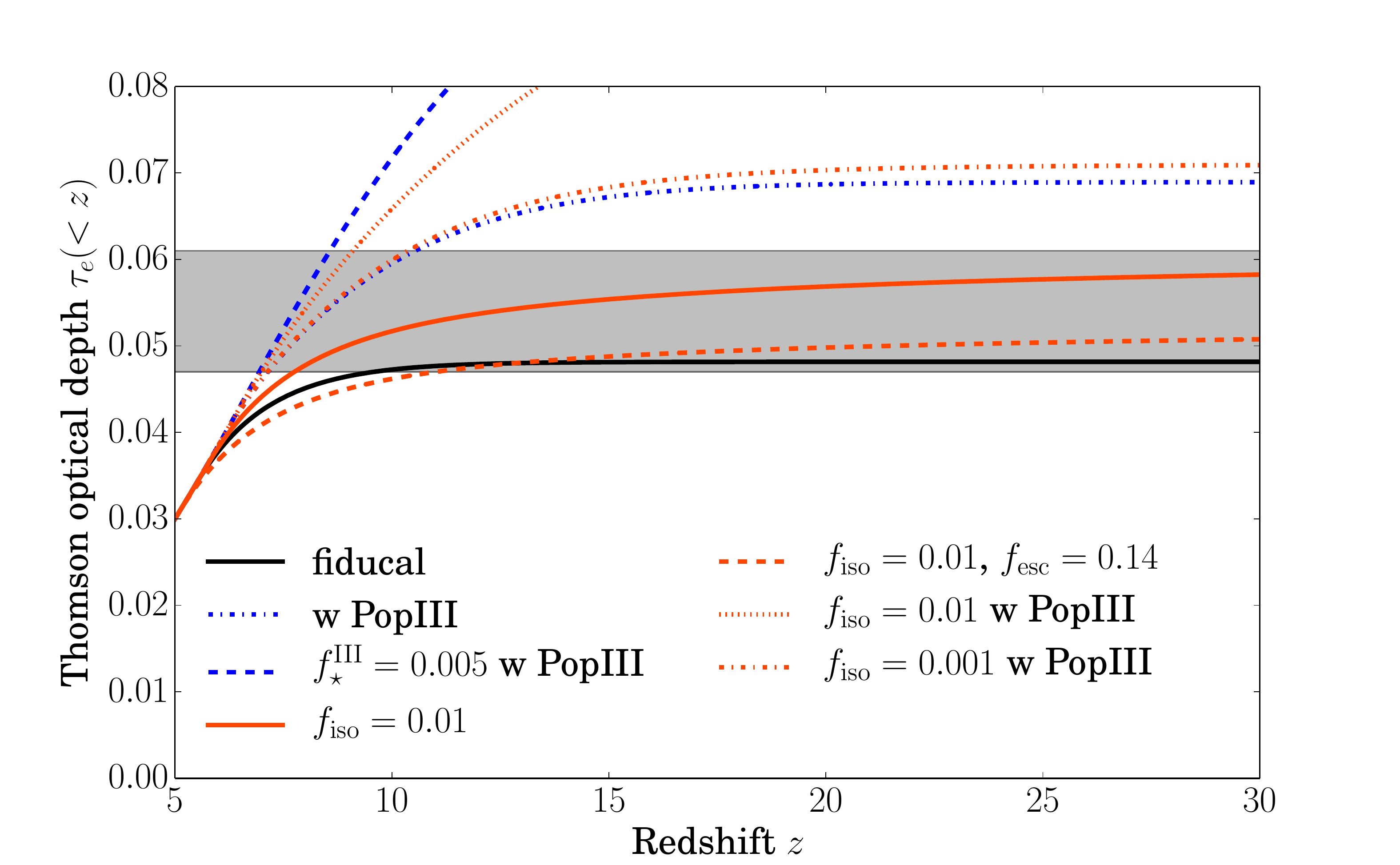}
\caption{Thomson optical depth to recombination, $\tau_e$. The grey band shows the Planck 2018 constraints on $\tau_e$. The models plotted are the same as in Fig.~\ref{fig_2}, with the black solid line respresenting the fiducial case using the stellar efficiency described in Sec.~\ref{glf_sec} and $f_{\rm iso} = 0$, while the orange lines show the contribution for varying $f_{\rm iso}$. The models that are clearly discrepant by the current CMB constraints have either (1) Pop-III photo-ionization and $f_{\rm iso}=0.01$ or (2) high Pop-III star formation efficiency and no isocurvature fluctuations (blue dashed). }
\label{fig_3}
\end{figure}

Another possibility of constraining the reionization process is utilizing the measurements of the CMB anisotropy, in particular the effect of the CMB Thomson scattering off of free electrons. Since the redshift where this would occur ($z<20$) are relatively closer than the surface of last scattering, this physical process affects predominantly large scales of the CMB fluctations. The CMB constraints from the Planck satellite on the $\tau_e$ are very strong \citep{2018arXiv180706209P}, as is show by the grey band in Fig.~\ref{fig_3}. The axion isocurvature model has a different signal in the Thomson scattering optical depth, which primarily reflects the prolonged redshift evolution of the reionization process seen in Fig.~\ref{fig_2}.  However, we note that that reionization affects on the CMB are not just as a single number, $\tau_e$, as an earlier tail ionization creates polarization anisotropies at smaller scales \citep{2003PhRvD..68b3001H, 2017PhRvD..95b3513H}.  An extended reionization is constrained by the Planck satellite to be $\tau_e(15, 30) < 0.007$, where this notation indicates the optical depth contributed between $z=15$ and $z=30$ \citep{2018arXiv180706209P}.  (The Planck limits on the tail of reionization vary only slightly with the assumed priors, and can lower the bound to $\tau(15, 30) < 0.006$ if flat priors are chosen on the positions of the knots on which $\tau$ is interpolated.)

\begin{figure}
\includegraphics[width=9cm]{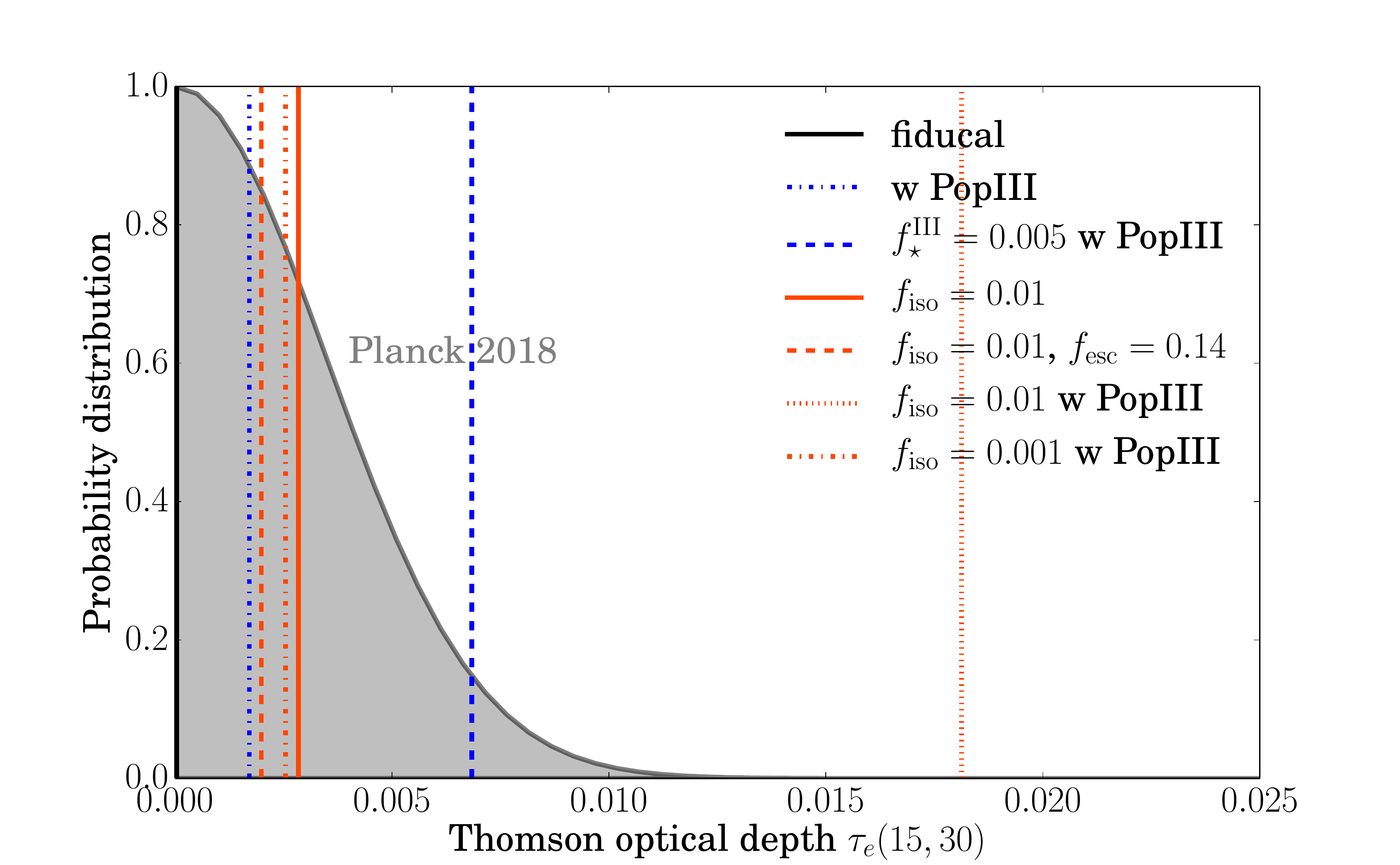}
\caption{Thomson optical depth contributed in redshift interval $15<z<30$, $\tau_e(15,30)$, in both models and observations. 
The grey posterior shows the Gaussian that yields the Planck 2018 $2\sigma$ upper bound of $\tau_e(15,30) < 0.007$. The vertical lines show $\tau_e(15,30)$ values for the same models that are plotted in Fig.~\ref{fig_3}: the fiducial case using the stellar efficiency described in Sec.~\ref{glf_sec} and $f_{\rm iso} = 0$ (black solid line), adding Pop-III photo-ionization rates (blue dot-dashed), and varying $f_{\rm iso}$ (orange lines with differing linestyles). Including even small Pop-III star formation efficiencies can result in detectable $\tau_e(15,30)$ for $f_{\rm iso}=0.01$. }
\label{fig_8}
\end{figure}

The limits on the tail of reionization are most constraining for models with an earlier star formation, in particular if the contribution of Pop-III stars is included. Of the models plotted in Fig.~\ref{fig_3} the models with $f_{\rm iso}=0.01$ and including Pop-III star formation is clearly excluded, with $\tau_e(15,30) = 0.018$ (dotted red line) as shown in Fig.~\ref{fig_8}. On the other hand, with the typical Pop-III star formation rate, the current data is not excluding a lower value of $f_{\rm iso}=0.001$, suggesting that lower $f_{\rm iso}$ values are more degenerate with astrophysical uncertainties of early star formation. Along this lines, increasing the star formation efficiency of Pop-III stars to $f_{\rm star}^{\rm III} = 0.005$ (0.05)\footnote{This is the efficiency one expects from assuming that each $10^5\;\mathrm{M_\odot}$ halo hosts one (ten) $100\;\mathrm{M_\odot}$ stars, and it further takes the efficiency to scale with halo mass.} leads to $\tau_e(15,30) = 0.007$ (0.017) for $f_{\rm iso}=0$. While tangential to the focus of this paper, this interestingly suggests that Planck is already constraining Pop-III star efficiencies in some of the range typically used. The limits on the tail of reionization are most constraining for models with an earlier star formation, in particular if the contribution of Pop-III stars is included. Apart from the stellar efficiency, changing the escape fraction of photons from Pop-II stellar population ($\tau_e(15,30)=0.002$ - dashed red line) can also lower the predicted optical depth, making isocurvature models similar to the fiducial adiabatic dark matter model (solid black line). This effect can also lower the optical depth in Pop-III models that have slightly higher $\tau_e$ compared to the CMB data (green and red dot-dashed lines).

The enhanced contribution to $\tau_e$ from the isocurvature fluctuations can be mimicked by astrophysical uncertainties: Similar effects can be observed by keeping $f_{\rm iso} = 0.01$ fixed, but switching off the Pop-III star formation (solid red line); or switching off isocurvature contribution, but adding Pop-III photo-ionization with the stellar efficiency of $f_{\star}^{\rm III} = 5\times 10^{-4}$ (green dot-dashed line). However, differences may show up in the tail of the reionization, where the aforementioned two models differ by a factor of $\approx 2$ in $\tau_e(15,30)$. In particular, further increasing Pop-III stellar efficiency by another order of magnitude to $f_\star^{\rm III} = 0.05$ results in too much ionization at early times -- $\tau_e(15,30) = 0.018$ -- which is ruled out by Planck CMB constraints. Such a high $\tau_e(15,30)$ is similar to that for the case with low Pop-III stellar efficiency and non-zero $f_{\rm iso}$ (see red dot-dashed line in Fig.~\ref{fig_8}). However, the contribution to the ionization fraction comes from $z<20$ in the case of high Pop-III stellar efficiency, while the signal in isocurvature models is dominated by the contribution at $z>20$.

On the other hand, further increasing the amount of isocurvature power by a factor of 5 ionizes the Universe to 10\% early on ($z\sim29 (46)$ for $f_{\rm iso} = 0.05 (0.1)$), leading to large values of $\tau_e(15,30) \sim 0.04 (0.09)$. Such models are clearly ruled out by the current CMB data, despite the astrophysical uncertainties. At a high enough level of $f_{\rm iso}$ the statement that such models are excluded by the CMB holds over the range of Pop-III efficiencies considered. In our models this transition happens in the range of $f_{\rm iso} = 0.01 - 0.1$.

Neglecting Pop-III contribution also lowers the effect of isocurvature modes. This occurs because the minimal mass ($M_{\rm min}$) that contributes to the Pop-II photo-ionization rates (Eqn.~\ref{eq:zeta_II}) is typically $\sim 10^8\;\mathrm{M_\odot}$, requiring a large $f_{\rm iso}$ to have an appreciable effect on these mass scales (see Fig.~\ref{fig_5}). On the other hand the minimal mass for Pop-III photo-ionization rates ($M_{\rm min}^{\rm III}$) is generally two orders of magnitude lower than for Pop-II stars ($\sim 10^6\;\mathrm{M_\odot}$), and thus more sensitive to smaller values of $f_{\rm iso}$.

Since some contribution from the Pop-III star formation is expected, values of $f_{\rm iso}$ of the order of $10^{-2}$ are excluded with the current measurements already, which corresponds to ALP mass limit of $m_a > 10^{-18}\;\mathrm{eV}$. Current and future CMB observations (e.g. CLASS, LiteBIRD) aim to put more stringent constraints on $\tau_e$ approaching the cosmic variance limit of $\sigma_\tau = 0.002$ \citep{DiValentino2018,Watts2019}. The sensitivity of measurements of the tail of reionization via statistics like $\tau_e(15,30)$ likely can be improved even more significantly over Planck with future missions than this improvement in $\sigma_\tau$  \citep{Watts2019}, although we expect measuring even higher redshift contributions like  $\tau_e(25,40)$ would be needed to be able to disentangle astrophysics and improve constraints on $f_{\rm iso}$.


Finally, we note that early ionization (which is likely also associated with X-ray and ultraviolet backgrounds) would shape the  high-redshift 21cm emission signal \citep{furlanetto-review}.  The 21cm signal is potentially sensitive to much lower star formation rate densities via these emissions than the ionizing emissions this section has focused on \citep{mcquinn12}.  The next section discusses another effect that may be even more constraining for this signal.

\section{CMB recombination and the dark ages thermal history}
\label{cmb_sec}
As illustrated in Fig~\ref{fig_7}, the presence of white noise isocurvature fluctuations leads to the formation of dark matter halos much earlier than in the standard scenario.  These early dark matter halos are moving supersonically relative to the gas, with an RMS Mach number of $\approx 2$ and with a Maxwellian distribution \cite{2010PhRvD..82h3520T}.  Some regions can even be moving hypersonically at $z\gtrsim500$ (i.e. with relative velocities of $\gtrsim 10$km~s$^{-1}$ so that the shocks can ionize the gas).  Furthermore, a $10^4 M_\odot$ dark matter halo will lose its velocity relative to the dark matter within a Hubble time \cite{oleary12}, potentially ionizing and heating the gas in the Universe if enough of these halos are present.

We first investigate the effect of shock ionization on the cosmic microwave background from such hypersonic motion.  Even percent level differences in the global $z\sim 500$ recombination history that result from this ionization could have a detectable effect on the cosmic microwave background \cite{2009PhRvD..80d3526S}.  However, while we found that the shocks in a large fraction of the Universe at $z>500$ would often heat the gas sufficiently for it to start to collisionally ionize, ionization would quickly sap out the thermal energy of the gas, leaving it at insufficient temperatures to collisionally ionize further. We found that because of this cost to ionization, even the strongest shocks would only ionize the gas to $\sim 1\%$.  This small ionization, coupled with the fact that (for viable $f_{\rm iso}$) only a fraction of dark matter has collapsed into the $M_h\gtrsim 10^3 M_\odot$ at $z\gtrsim 500$ halos that generate significant shocks, results in the recombination history being negligibly affected.

We next turn to the heating imparted by such shocks.  If the heating occurs early enough, it could also affect the recombination history, as the recombination rate depends inversely on the temperature.  Our calculations suggest that such heating does not occur at early enough times to be relevant for Recombination.  Another observable is the cosmological 21cm signal. When the 21cm signal is in absorption as is anticipated $15\lesssim z 
\lesssim 30$, its amplitude is inversely proportional to the gas temperature \citep{furlanetto-review}.   We show below that this  shock heating could be important for this 21cm signal.

\begin{figure}
\includegraphics[width=8cm]{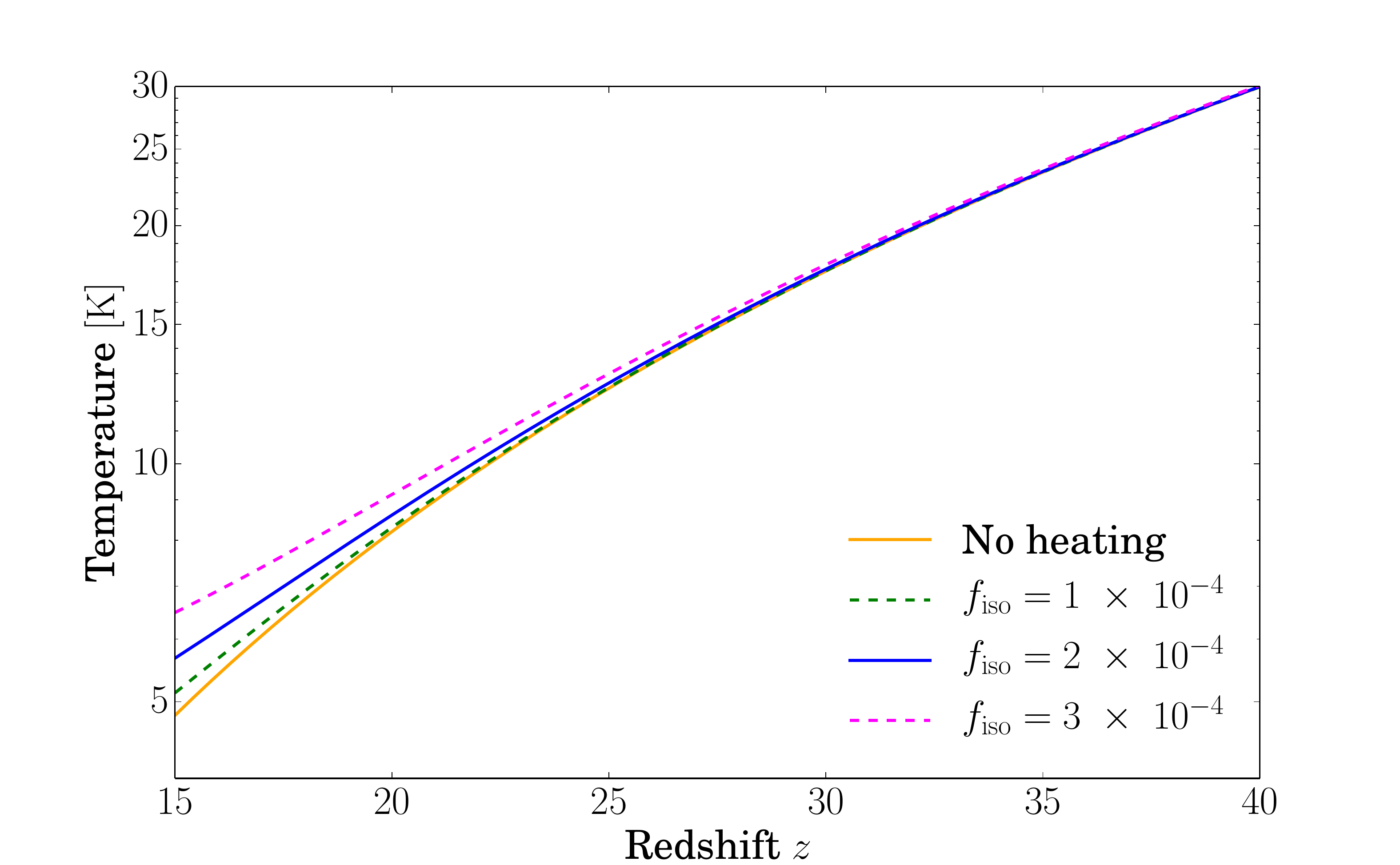}
\caption{Rough estimates for the evolution of gas temperature at different $f_{\text{iso}}$, with larger $f_{\text{iso}}$ leading to more shock heating. Our estimates suggest that $f_{\text{iso}} \sim 10^{-4}$ lead to percent-level or greater shock heating. Percent level heating would manifest in a qualitatively different high-redshift 21cm signal, with significant BAO peaks.}
\label{fig_6}
\end{figure}

A simple estimate for the amount of shocking uses that we know how much energy is dissipated into the gas via dynamical friction, a frictional force from the gas that acts to decelerate the supersonicly streaming dark matter halos.  Namely, halos more massive than $\sim 10^5-10^6 M_\odot$ should lose all of their relative velocity to the baryons in a Hubble time at $z\sim 20$ \citep{mcquinn12}.  Some of this dynamical energy should go into shocks (and if all of the energy goes into shocks we would expect to heat the Universe by $\langle {\cal M}^2 \rangle  \sim 4$).
We estimate the effect of shock heating on the thermal history by solving
\begin{eqnarray}
    \frac{dT}{dt}&=&\overbrace{-2HT}^{\rm adiabatic}+ \overbrace{\frac{8\pi^2 x_i T_{\gamma}^4\sigma_T(T_{\gamma}-T)}{45m_e(1+x_i)}}^{\rm Compton} \label{eqn:shockheating}\\
    &-&\underbrace{ \frac{ \mu m_p}{3 M_P^4} \langle \zeta_s(v_{\text{b-dm}}) v_{\text{b-dm}}^{-1} \rangle \int_{M_{\rm min}}^{M_{\rm max}} dM_h M_h^2 \frac{dn}{dM_h}}_{\rm shock~ heating},   \nonumber 
\end{eqnarray}
where $\sigma_T$ is the Thomson cross section, $m_p$ is the mass of hydrogen atom, $v_{\text{b-dm}}$ is the velocity difference between dark matter and baryons, $M_h$ is the halo mass and $\rho_{\text{dm}}$ is the density of dark matter. The Compton cooling term owes to the scattering of CMB photons, which is negligible below redshift $z=200$. The ``shock heating'' term in Eqn.~\ref{eqn:shockheating} follows from the power generated from dynamical friction; taking the expression in \cite{1999ApJ...513..252O} but dropping the factor of the Coulomb logarithm.  The motivation for dropping this logarithm is that the resulting expression accounts only for gas that intersects within the Bondi-Hoyle radius for accretion ($r_{\rm BH} = 2GM/v_{\text{b-dm}}^2$; \citep{1944MNRAS.104..273B} and see this endnote \footnote{Our expression for the heating power from each halo is equal to the cross section for Bondi-Hoyle accretion times the kinetic energy density of the accreted gas times the velocity offset. }), which is the gas whose trajectory would be deflected to the origin (in the absence of pressure) and hence is most likely to shock.  We conservatively assume the shock heating has efficiency $\zeta_s$ at thermalizing its energy, and we take $\zeta_s=0.1$ motivated by entropy increase calculated in planar shocks with Mach numbers of ${\cal M}=2$.  Finally, $M_{\rm max}$ is set to the halo mass whose timescale to lose its energy by dynamical friction in much less than the age of the Universe, as once a halo reaches this mass, it will likely have decelerated and no longer contribute to the heating.  We take $10^6 M_{\odot}$ as the maximum mass. The minimum mass is set by where the halo viral radius equals $r_{\rm BH}$, which we find is $M \approx 10^4 M_{\odot}$. It is worth stressing that the shock heating effect is most sensitive to the maximum mass. If we make the maximum mass a factor of 10 smaller ($10^5 M_{\odot}$), the temperature difference will be about three times smaller in Fig.~\ref{fig_6}, which we think reflects the level of uncertainty.

Our simple estimates show that the shock heating effects from axion halos starts to become significant around redshift $z=20$ as shown in Fig.~\ref{fig_6} for $f_{\rm iso}\gtrsim 10^{-4}$. Models predict a global 21cm absorption feature at $\sim 80$MHz, corresponding to absorption at $z\sim 15-20$ \cite{furlanetto-review}, the same signal purported to be detected by EDGES \citep{2018Natur.555...67B}.  This absorption dip is inversely proportional to the gas temperature.  Thus, a detection of the full amplitude of this dip should at a minimum be used to discern shock heating at the ${\cal O}(1)$ level, requiring for us $f_{\rm iso}\sim 5\times10^{-4}$.  Such heating would be hard to disentangle from X-ray heating from the first supernovae and black holes \citep{furlanetto-review}. However, efforts to detect fluctuations in the 21cm have a potentially smoking gun signal for this heating.  Since change in temperature is tied to the relative velocity between the baryons and the dark matter ($v_{\text{b-dm}}$), and this relative velocity is modulated by the acoustic physics in the early Universe, any heating could result in large acoustic oscillations in the signal. \citet{mcquinn12} showed that even just $\sim 3\%$ changes in the temperature that are tied to $v_{\text{b-dm}}$ would lead to order-unity acoustic features in the 21cm signal at $k\sim 0.1$Mpc$^{-1}$ \citep{mcquinn12}, qualitatively changing the 21cm signal.  Our estimates in Fig.~\ref{fig_6} suggests that heating at the few percent level occurs for $f_{\rm iso}\gtrsim 10^{-4}$, although we illustrate the rough constraint in Fig~\ref{fig_constraints} at $f_{\rm iso}=3\times 10^{-4}$.  These acoustic features are quite distinct from the smoother continuum of fluctuations from the extra star formation would create, which were referenced as a potential observable in \S~\ref{sfr_sec}.

\section{Conclusions}
\label{conclude_sec}

One possible candidate for the dark matter is that it is an ultra-light scalar field that is generated in the early universe in a similar manner to that for the QCD axion, making it an `axion like particle' (ALP). Most previous studies have concentrated on how the ultralight ALP's quantum pressure suppresses the small-scale growth of the adiabatic fluctuations from inflation or on how its relaxation can lead to solitonic cores \citep{Schive2014,Veltmaat2018,Mocz2019}. However, if the symmetry breaking that establishes the axion-like particle (ALP) occurs after inflation ends, this leads to white isocurvature fluctuations in the ALP energy density.  The parameter space where the post inflationary scenario can occur are for symmetry breaking scales of $10^{13}-10^{16}$GeV for the particle mass ranges that are probed by the large scale structure observables considered here ($m_A\sim 10^{-13}-10^{-20}$eV).  The higher values for the symmetry breaking scale (and lower values for the mass) push against limits from searches for inflationary $B$-modes. This paper focused on how these isocurvature fluctuations could influence various observations of early structure formation.

Fig.~\ref{fig_4} summarizes our resulting constraints on the ALP mass $m_a$ and isocurvature fluctuation amplitude $f_{\rm iso}$ -- defined in the traditional manner as their ratio with adiabatic fluctuations at a wavenumber of $0.05$Mpc$^{-1}$ (but we also report constraints in terms of the more natural nonlinear wavenumber $k_{\rm nl}$). The solid lines show the relation between the axion mass and $f_{\rm iso}$.  Different colours represent different parameterizations of the evolution of the axion mass with temperature after it commences oscillations. The simplest model, and also most conservative in terms of mass constraints, is the $n=0$ case where the mass was set at early times. For an ALP coupled to an asymptotically free sector (in analogy to the QCD axion), leading to a mass that increases in size until the cosmic temperature falls below the sector's confinement scale, the value of $n$ is nonzero (with $n = 4$ approximating the evolution of the QCD axion). As $n$ increases above $4$, the sensitivity of our results to $n$ becomes weak.

The cosmological observables presented in this paper are sensitive to different axion masses $m_A$ or equivalently different levels of $f_{\rm iso}$, with the smaller scale the observable is sensitive to the stronger the constraint. Our strongest present constraint comes from the Ly$\alpha$ forest power spectrum measurements at high redshifts (orange dashed line).  The lower bound on the ALP mass from the Ly$\alpha$ forest is $m_A > 2\times 10^{-17}\;\mathrm{eV}$ for $n=0$ (and $m_A > 10^{-13}\;\mathrm{eV}$ for $n=4$). Apart from being currently most constraining bound, the Ly$\alpha$ analysis is also the least affected by uncertainties in the astrophysics of the existing probes we investigated.

Another potential probe is high-redshift galaxy observations. We find that only for $m_A$ already ruled out by the Ly$\alpha$ forest is the observed luminosity function  qualitatively changed in a manner that could potentially be disentangled from more mundane astrophysical explanations.  However, smaller mass (and higher redshift) galaxies than can be observed directly are more substantially boosted by isocurvature fluctuations.  Such diminutive galaxies may be observable via their effect on the ionized fraction evolution during the Reionization Epoch.  We find that a particularly interesting observable is the CMB, which is sensitive to the high-redshift tail of reionization.  This tail can be substantially more extended in models with white isocurvature fluctuations. While we find that the ionization fraction in models where galaxies form via the traditional route (in halos massive enough that the gas can cool atomically) only show qualitatively different trends for $m_A$ already ruled out by the forest, models that include Pop-III stars (even for much lower efficiencies for their formation than is commonly assumed) could lead to a small residual ionization to extremely high redshifts.  Thus, future CMB efforts could potentially probe $m_A$ range similar to that of the Ly$\alpha$ forest. 

Finally, the shock-heating of the gas due to supersonically moving axion minihalos during the Cosmic Dark Ages and Cosmic Dawn could lead to even stronger constraints, potentially excluding ALP masses of $m_A < 10^{-16}\;\mathrm{eV}$ for $n=0$. This shocking would suppress the depth of the absorption trough in the global 21cm signal (as probed by e.g EDGES and PRIZM). The caveat is that X-ray heating could have a similar effect \citep{Barkana2018,Fialkov2019}. However, even percent-level changes in the mean temperature from shock-heating will manifest in distinct baryon acoustic oscillation features in the 21cm brightness temperature fluctuations that trace the relative baryon-dark matter velocity field.  These oscillations are potentially a smoking gun of shock heating from a dramatic enhancement in the number of minihalos.

Some low redshift small-scale structure probes could complement the probes discussed here. First, local observations of Milky Way tidal streams could lead to detection of small sub halos in the mass range $10^8 - 10^5\;\mathrm{M_\odot}$ \citep{Bovy2017, Bonaca2019}, with some uncertainty in whether the lowest values of $10^5\;\mathrm{M_\odot}$ can be disentangled from astrophysical uncertainties, as encounters with these subhalos open up gaps in these streams.  This places the sensitivity of the galactic streams somewhere in the range of isocurvature amplitudes of $f_{\rm iso} = 0.001 - 0.01$, potentially pushing the constraints lower than the current Ly$\alpha$ bound and comparable to our most optimistic  reionization constraints. 

In addition, \citet{Dai2019} recently showed that the micro-lensing caustics of stars on a cluster macro-lens could even be sensitive to the minute value of $M_{H(m_A)}$ for the QCD axion of $\sim 10^{-12}M_\odot$, where $M_{H(m_A)}$ is the mass within the Horizon at $T_{\rm osc}$.  In particular, these micro-lensing caustics are perturbed by these axion structures, deviating from the smooth profile otherwise expected. This constraint can also be translated to our scenario. \citet{Dai2019} showed this method is sensitive to $10^{-13} <M_{H(m_A)}  < 10^{-6}\;\mathrm{M_\odot}$, which translates to the bounds on the ALP mass of $10^{-15} < m_A < 10^{-11}\;\mathrm{eV}$  for $n=0$ ($10^{-11} - 10^{-6}\;\mathrm{eV}$ for $n=4$). Since the sensitivity falls off on both sides of the ALP mass range, this makes the microlensing of stars complementary to the signatures of early structure formation considered in this paper.  Future observations with HST or JWST should be able to push forward this exciting science \citep{Chen2019,Kaurov2019}.

Lastly, a post-inflation ALP may affect the properties of black holes. Studies of black hole superradiance \citep{Arvanitaki2015,Baryakhtar2017,Scott2018,Davoudiasl2019} -- the gravitaional production of a ALP halo from the free energy in black hole spin -- have excluded the existence of ALPs with $10^{-14} < m_A < 10^{-11}\;\mathrm{eV}$ from measurements of finite stellar black hole spins. The measurements of super massive black hole spin can potentially exclude a wide mass range $m_A < 10^{-16}\;\mathrm{eV}$ \citep{Scott2018}, but inferring the black hole masses over a broad mass range.  The bounds from superradiance are also only valid in the limit of $f_A > 10^{14}\;\mathrm{GeV}$ and no self-interaction \citep{Scott2018}. Furthermore, the earlier structure formation sourced by a post-inflation ALP could potentially produce the seeds that grow into the highest mass black holes, ameliorating somewhat the difficulty in having sufficient time for these seeds to grow into the highest redshift quasars (e.g. \citep{Latif2016}).

\appendix
\renewcommand\thefigure{\thesection.\arabic{figure}}
\setcounter{figure}{0}

\bigskip

\acknowledgments We would like to thank Akshay Ghalsasi for helpful conversations, and Erik Anson for running tests of the universality of the mass function in cosmologies near our white case. VI and MM thank US NSF grant AST-1514734, and MM and HX the University of Washington Royalty Research Grant program. HX is also supported in part by the U.S. Department of Energy, under grant number DE-SC0011637. VI acknowledges support by the Kavli Foundation.

\bibliography{references}

\end{document}